\documentclass[%
 pre,
 amsmath,amssymb,
 reprint,%
floatfix,
superscriptaddress
]{revtex4-1}

\usepackage{graphicx}% Include figure files
\usepackage{dcolumn}% Align table columns on decimal point
\usepackage{bm}% bold math
\usepackage{soul}
\usepackage[utf8]{inputenc}
\usepackage[T1]{fontenc}
\usepackage{mathptmx}
\usepackage{etoolbox}
\usepackage{amsmath}
\usepackage{xcolor,hyperref}
\hypersetup{
   colorlinks,
   linkcolor={blue!50!black},%{red!80!black},
   citecolor={blue!50!black},
   urlcolor={blue!80!black}
}
\makeatletter
\def\@email#1#2{%
 \endgroup
 \patchcmd{\titleblock@produce}
  {\frontmatter@RRAPformat}
  {\frontmatter@RRAPformat{\produce@RRAP{*#1\href{mailto:#2}{#2}}}\frontmatter@RRAPformat}
  {}{}
}%
\makeatother

\draft % marks overfull lines with a black rule on the right

\begin{document}

%\title{Plasticity onset in cyclically sheared amorphous systems with varying compressibility} 
\title{Finite compressibility and strain hardening in elasto-plastic models of amorphous matter.}
%Title of paper

\author{A.~Elgailani}
\affiliation{Department of Mechanical and Industrial Engineering,
Northeastern University, Boston, Massachusetts 02115, USA}

\author{D.~Vandembroucq}
\affiliation{PMMH, CNRS UMR 7636, ESPCI Paris, PSL University, Sorbonne  Université, Université Paris Cité, F-75005 Paris, France}

\author{C.E. ~Maloney}
\affiliation{Department of Mechanical and Industrial Engineering, Northeastern University, Boston, Massachusetts 02115, USA}

\date{\today}

\begin{abstract}
We study a mesoscopic elasto-plastic model of amorphous matter with varying dimensionless compression modulus, $K/\mu$, where $K$ and $\mu$ are the compression and shear moduli.
We study both cyclic shear with amplitude $\Gamma$ and forward steady shear.
In cyclic shear, the terminal behavior is, in order of increasing $\Gamma$: i) trivially elastic, ii) hysteretic but with microscopically reversible limit cycles, iii) diffusive with no return to previously visited configurations.
%In previous work, for the incompressible case, we found that the transition between i) and ii) at the onset point $\Gamma_0$, was determined by the so-called Eshelby back stress, $\sigma_0$.
%Here we show that precisely the same results follow after we account for the dependence of $\sigma_0$ on the finite value of $K/\mu$.
We show that the the transition between i) and ii) at the onset point $\Gamma_0$ is determined by the so-called Eshelby back stress, $\sigma_0$, which depends on the Poisson ratio.
The result is that systems which are more compressible, with smaller $K/\mu$, are effectively harder with a higher value of $\Gamma_0$ and a correspondingly larger regime of purely elastic behavior in cyclic loading.
In forward shear, we show that $\sigma_0$ plays a similar role where lower $K/\mu$ results in a higher value of the steady state flow stress, $\sigma_y$.
%The dependence of $\sigma_y$ on $K/\mu$ is slightly different for $\sigma_y$ than for $\Gamma_0$, and we explain this in terms of the distribution of local stresses, $P(\sigma)$.
We show how increasing $K/\mu$ results in an increase in the amplitude of the stress \emph{redistribution} after a local yielding event without resulting in a change in the net stress \emph{relaxation} from the event and discuss how this is related to the assumptions which go into mean-field descriptions of amorphous solids.
Another striking feature of the model is the emergence of a complex hardening behavior in the absence of any microspcopic ad-hoc hardening parameters or rules. 
In particular, we observe a transition between a kinematic and an isotropic hardening behavior precisely at the shear cycling amplitude $\Gamma_0$ associated with the hysteresis transition.
The enhanced plastic response for incompressible systems is also seen in amorphous alloys where it is usually attributed to excess free volume, while in the present model, it arises simply as a consequence of the dependence of the Eshelby backstress on the Poisson ratio.
Our results should have important implications for amorphous metallic alloys or other glassy systems (such as colloidal glasses, emulsions, etc.) where $K/\mu$ can vary with composition, age, quench procedure, or mechanical processing history.

% In cyclic shear, below a certain threshold, $\Gamma_0$, after a transient, the system becomes completely elastic with no plasticity or energy dissipation, while above $\Gamma_0$, the final steady state exhibits plasticity with hysteretic energy dissipation.
% Here we study how $\Gamma_0$ depends on $K$ with the shear modulus, $\mu$, fixed. 
% As in the incompressible case, we find that $\Gamma_0$ is set by the Eshelby back stress, $\sigma_0$, with $\Gamma_0=(1+\sigma_0)/2$.
% $\sigma_0$, in turn, decreases monotonically with increasing $K/\mu$.
% Accordingly, incompressible systems have smaller $\Gamma_0$ than systems with finite $K/\mu$. 
% Furthermore, we show a qualitatively similar effect of $K/\mu$ on the flow stress, $\sigma_y$, in steady forward shearing, but less pronounced than on the $\Gamma_0$ value obtained in cycling.
% This effect on $\sigma_y$ can also be understood in terms of the $K/\mu$ dependence of the back stress.
% Our results should have important implications for amorphous metallic alloys and glassy colloidal systems where $K/\mu$ can vary with composition or age of the glass.
\end{abstract}

\maketitle %\maketitle must follow title, authors, abstract and \pacs
\section{Introduction}
%Amorphous systems at different scales, such as metallic glasses, foams, emulsions, bubble rafts, colloids, pastes, and dense granular packings, while at the level of individual particles are quite different, exhibit similar phenomenology.
%At sufficiently low strains, they respond elastically like solids where their microscopic configuration is conserved while at sufficiently high amplitudes they flow and microscopic rearrangements of particles take place continuously,,,,,,,ISI:000337136700009,ISI:000305924900009
The plastic response and yielding behavior of amorphous solids such as amorphous alloys, soft glasses (colloidal glasses, emulsions, pastes, foams, etc.), and dense granular matter is understood to be controlled by local shear transformations~\cite{ARGON197947,Falk:1998aa,Rodney_2011,picard2004elastic,PhysRevE.74.016118,Falk:2010aa,PhysRevE.90.042305,nicolas2018deformation}.
After a local shear transformation, the surrounding material is subjected to stress increments as in Eshelby's inclusion problem~\cite{doi:10.1098/rspa.1957.0133,Weinberger2005LectureN,picard2004elastic,Mura:1982ud}.
It is a well known result from Eshelby's solution to the inclusion problem that after a transformation, an inclusion will find itself under a stress that opposes the transformation just incurred~\cite{ARGON197947,Schuh:2007wv,Mura:1982ud,Weinberger2005LectureN}.
In Eshelby's solution, the entire stress field in general, and the stress on the inclusion in particular,  will depend on the compressibility of the material, typically expressed in terms of the Poisson ratio, $\nu$, or the ratio of compression modulus to shear modulus, $K/\mu$.
The higher the value of $K/\mu$, the higher the resulting stress both in the inclusion and in the surrounding matrix for a given amplitude of plastic strain in the inclusion.
This effect of finite compressibility is accounted for in Argon's seminal work on shear transformations~\cite{ARGON197947,Schuh:2007wv} where the energy barrier for thermally activated stress-biased shear transformations depends quantitatively on the Poisson ratio.
However, somewhat surprisingly, quantitative accounting for finite compressibility is missing from most recent treatments of amorphous matter based on local shear transformations.

At the same time, it is well known that the Poisson ratio is empirically related to the ductility of amorphous alloys.
In general, metallic glasses with high $K/\mu$ are more ductile with higher fracture toughness~\cite{Schuh:2007wv}.  
Various authors interpret the result in slightly different terms~\cite{Lewandowski-:2005ua,Schroers2004DuctileBM,Wang-JAP06,Rouxel-NatMat11}, but broadly speaking, a high value of $K/\mu$ indicates that it is relatively easy to trigger shear-induced plastic relaxation and relatively difficult to cavitate and nucleate fracture.
One might also expect the low values of $\mu/K$ in ductile systems to be related to excess low-lying soft modes in the system both associated with  reduction in \emph{shear} modulus and acting as modes for plastic relaxation~\cite{PhysRevMaterials.4.113609,PhysRevE.97.033001,PhysRevLett.117.045501,PhysRevE.101.032130,doi:10.1073/pnas.1919958117}.
However, a quantitative understanding of this relationship between $K/\mu$ and ductility is lacking.

In this work, using an elasto-plastic automaton based on a space-filling tiling of identical potential shear transformation sites, we show how the increasingly negative Eshelby back stress, $\sigma_0$, and the associated larger redistribution of stresses at increasing values of $K/\mu$ allows systems to yield earlier at lower stresses and strains both in cyclic shear and in steady forward shear.
In~\cite{elgailani2022anomalous}, we studied the cycling amplitude, $\Gamma_0$, required to observe plastic hysteresis in the terminal steady states for an incompressible system. 
Here, in the compressible systems, we show we show that while $\sigma_0$ depends on $K/\mu$, the previous $\Gamma_0$ dependence on $\sigma_0$ is in complete agreement with our previous work~\cite{elgailani2022anomalous}.
In the case of forward steady shear, we must slightly modify our scaling arguments based on $\sigma_0$, but we arrive at a prediction for $\sigma_y$ that differs quantitatively from the prediction for $\Gamma_0$, 
%but 
although it has the same overall trend.
%In the former case of cyclic shear, the connection is perhaps clearest and most precise, and characterization of the role of finite compressibility in cyclic shear will serve as the main focus of the present work.

Amorphous systems subject to forward-reverse shear cycling have been the focus of an increasing number of experimental, theoretical, and numerical studies over the past several years~\cite{lundberg2008reversible,Knowlton-SM2014,C3SM51014J,PhysRevLett.112.028302,PhysRevResearch.2.012004,Keim:wi,Keim:vj,Keim:2019vm,doi:10.1063/5.0085064,fiocco2013oscillatory,  Leishangthem:2017us, doi:10.1073/pnas.2100227118,priezjev2013heterogeneous,Priezjev:2016wq,Szulc:2022uc,Galloway:2020ug,regev2013onset,Regev:2019ut,Regev:2015wt,Regev-PRE20-Annealing-Cycle,PhysRevE.103.062614,Mungan-PRL19,https://doi.org/10.48550/arxiv.2211.03775,Khirallah-PRL2021,doi:10.1063/5.0102669}.
In their seminal work \cite{Keim:vj}, Keim and coworkers identified three steady state response regimes for an amorphous material under cyclic shearing: at low amplitudes, the material eventually becomes elastic with no rearrangement of particles or atoms. 
At intermediate amplitudes, the material locks into limit cycles~\cite{regev2013onset} with energy dissipating plastic rearrangements that precisely reverse themselves after an integer number of strain cycles to bring the system back to a previously visited microscopic configuration.
At large amplitudes, the particle trajectories are diffusive, and the system never returns to any previous microscopic configuration.
These three regimes define two transitions: the \emph{hysteresis} transition from the elastic regime (ER) to the reversible plastic regime (RPR) at the amplitude $\Gamma_0$, and the \emph{irreversibility} transition from the RPR to the diffusive regime (DR) at the amplitude $\Gamma_y$ where $\Gamma_y>\Gamma_0$.

Most works in the literature on forward-reverse cycling have focused on the later irreversibility transition at $\Gamma_y$, while the earlier hysteresis transition at $\Gamma_0$ has only been studied more recently.
In reference~\cite{Khirallah-PRL2021}, some of us studied an automaton elastoplastic model---based on the idea of tiling space with sites any of which could undergo a discrete plastic transformation according to the Eshelby theory \cite{Weinberger2005LectureN}--- and reproduced the three regimes. 
%They showed that in approaching the RPR->DR transition point from below the number of cycles the system requires to reach a limit cycle diverges while from above, the diffusion coefficient vanishes. 
%They further showed that the average period increases as they approach the transition point from below.
That work focused on the later transition, the irreversibility transition, at $\Gamma_y$.
Later, in reference~\cite{elgailani2022anomalous}, we focused on the earlier transition, the hysteresis transition, at $\Gamma_0$. 
We showed that while increasing the strain amplitude in the ER results in a harder terminal state with lower energy (analogous to the case with thermal annealing), increasing the strain amplitude in the RPR results in a softer material, with a lower plasticity onset strain $\gamma_*$, despite the fact that the energy continues to decrease with increasing amplitude.
In our model, there is a characteristic value of stress, $\sigma_0$, that a site will find itself under after it reaches the plastic threshold and transform; a so-called back stress.
We found that $\sigma_0$ manifests itself as a characteristic feature in the stress distribution that determines $\gamma_*$ and its dependence on $\Gamma$.
We showed that $\Gamma_0$ is linked to $\sigma_0$ by the relation: $\Gamma_0=\frac{1+\sigma_0/\mu}{2}$.  
We finally showed that $\Gamma_0$ corresponded to a state that is optimally hardened. 

Here we extend these results to the case of finite compressibility. 
We discuss both oscillatory shear driving and forward shear driving.
For the parts of this work concerning shear cycling, we continue to focus on the hysteresis transition at $\Gamma_0$ using the same model as in reference \cite{elgailani2022anomalous}, but here, we allow for a finite compression modulus and investigate its effect.
The only adjustable material parameter in our study is the dimensionless compressibility,  $K/\mu$, where $K$ is the compression modulus and $\mu$ is the shear modulus. 
We find that $\Gamma_0$ increases as we decrease $K/\mu$. 
Apart from that, everything we found in our previous study \cite{elgailani2022anomalous} continues to hold for a material with finite compressibility.
For instance: 1) We show that the relation $\Gamma_0=\frac{1+\sigma_0/\mu}{2}$ is maintained. 
However, now $\sigma_0$ depends on $K/\mu$. 
2) The plastic strain, $\epsilon$, vs. the total strain, $\gamma$, curves have the same shape in \cite{elgailani2022anomalous} where above $\Gamma_0$ the plastic strain rate, $\frac{d\epsilon}{d\gamma}$, jumps to a finite value of roughly $0.2$ at $\gamma_*$ independent of $\Gamma$ or $K/\mu$.
3) Below $\Gamma_0$, the dissipated energy and the plastic strain are precisely zero but above $\Gamma$ they continuously increase with $\Gamma$ from zero with the plastic strain amplitude $\epsilon_p$ following a powerlaw with $\epsilon_p\propto (\Gamma-\Gamma_0)^{1.2\pm0.1}$ 
4) Both $\gamma_*$ and the stress value at $\gamma_*$, $\sigma_*$, decrease with $\Gamma$ indicating a softening effect beyond $\Gamma_0$ and the relation $\gamma_*=\Gamma_0-(\Gamma-\Gamma_0)$ is conserved.
5) $\sigma_0$ gives rise to a characteristic, $\Gamma$ independent feature in the stress distribution $P(\sigma)$ that explains the nontrivial dependence of $\gamma_*$, $\sigma_*$ and $\epsilon_p$ on $\Gamma$.

For forward shear, the situation is slightly more subtle.
We show that for all $K/\mu$, the $P(\sigma)$ distributions in steady state show a pronounced shoulder at $\sigma_0$.
After rescaling the distributions to take into account the $K/\mu$ dependent $\sigma_0$, they show an almost perfect collapse.
This allows us to express the dependence of the average flow stress in steady state, $\sigma_y$, in terms of $\sigma_0$.
This dependence of $\sigma_y$ on $\sigma_0$ turns out to be qualitatively similar to but quantitatively different from the $\Gamma_0$ dependence on $\sigma_0$.
Both $\sigma_y$ and $\Gamma_0$ decrease with increasing $K/\mu$.

%In addition to cyclic shear, we also study steady forward shear and show a similar trend for the flow stress, $\sigma_y$, with $\sigma_y$ decreasing as $K$ increases.
%Furthermore, we observe a similar feature in the $P(\sigma)$ distribution at $\sigma_0$ even in forward shear.

% We find that this dependency of $\sigma_0$ on $K/\mu$ agrees with the continuum solution where the ratio $\frac{\sigma_0}{\sigma_{0cir}}$ approaches a constant value $0.7268$  as the discretized system size, $L$, approaches infinity. 

In the following we first discuss in section~\ref{sec:eshelby} the effect of finite compressibility on the Eshelby stress field induced by a plastic inclusion and detail the numerical discretization of the linear elasticity equations and stress equilibrium. 
In section~\ref{sec:model} we present the mesoscopic elastoplastic model. 
Then we discuss the effect of finite compressibility on cyclic loading behavior in section~\ref{sec:cyclic-loading}. 
In particular we show that the material becomes increasingly ductile as the compressibility decreases -- with the incompressible state being maximally ductile -- and relate this trend to the dependence of the stress redistribution kernel on $K/\mu$. 
In section~\ref{sec:forward-shear} we discuss the forward shear loading behavior and analyze the decrease of the flow stress with decreasing compressibility; again indicating that the incompressible state is maximally ductile. 
In section~\ref{sec:discussion},  we explain the emergent hardening behavior in terms of simple schematic hardening model. 
We finally summarize and discuss our main results in section~\ref{sec:conclusion}.

\section{Eshelby inclusions: Effect of compressibility on continuum solution and staggered grid discretization}
\label{sec:eshelby}

\subsection{Continuum solution}
When an inclusion inside an initially stress free material undergoes a plastic transformation with a plastic strain $\epsilon_{ij}$, known as the eigenstrain, the inclusion and the surrounding material will exert equal and opposite forces against each other.
%The Eshelby problem is to find the stress $\sigma_{ij}$, displacement gradient $\gamma_{ij}$, and displacement $u_i$ in both the transformed inclusion and the matrix. 
The Eshelby problem is to find the resulting displacement, $u_i$, its gradient, the linear strain, $\gamma_{ij}=\partial_i u_j$, and the resulting stress, $\sigma_{ij}$ in both the transformed inclusion and the matrix. 
For an elliptical inclusion with homogeneous plastic strain, Eshelby showed that the total strain (also the stress) inside the inclusion is also spatially homogeneous.
The Eshelby tensor, $S_{ijkl}$, relates the total strain inside the inclusion to the prescribed plastic strain: 
\begin{equation}
\gamma_{ij}=S_{ijkl}\epsilon_{kl}\;.
\label{eq:Eshelby_tensor}
\end{equation}
The solution for a circular inclusion in an infinite linear elastic medium is provided in \cite{Weinberger2005LectureN} as:
% \begin{equation}
%     \mathbf{S}_{ijkl}=\frac{4\nu-1}{8(1-\nu)}\delta_{ij}\delta_{kl}+\frac{3-4\nu}{8(1-\nu)}(\delta_{ik}\delta_{jl}+\delta_{il}\delta_{jk})
% \end{equation}
\begin{equation}
    S_{ijkl}=\frac{\lambda-\mu}{4\lambda+8\mu}\delta_{ij}\delta_{kl}+\frac{\lambda+3\mu}{4\lambda+8\mu}(\delta_{ik}\delta_{jl}+\delta_{il}\delta_{jk})\;,
\label{eq:eshelby_tensor_circle-lameS}
\end{equation}
where the linear elastic material has an energy density, $\frac{1}{2}C_{ijkl}(\gamma-\epsilon)_{ij}(\gamma-\epsilon)_{kl}$, and the isotropic $C$ tensor is described by the shear modulus, $\mu$ and Lame parameter, $\lambda$: $C_{ijkl}=\lambda\delta_{ij}\delta_{kl}+\mu(\delta_{ik}\delta_{jl}+\delta_{il}\delta_{jk})$. 
Note the compression modulus, $K$, in 2D is $K=\lambda+\mu$.
In this article, we use the three components of Voigt notation for $\gamma$, and $\epsilon$, tensors referring to them as modes 1, 2, and 3 defined as:
\begin{eqnarray}
\gamma_1&=&\gamma_{xx}-\gamma_{yy}\;,\\
\gamma_2&=&\gamma_{xy}+\gamma_{yx}\;,\\
\gamma_3&=&\gamma_{xx}+\gamma_{yy}\;,\\
\epsilon_1&=&\epsilon_{xx}-\epsilon_{yy}\;\\
\epsilon_2&=&\epsilon_{xy}+\epsilon_{yx}\;,\\
\epsilon_3&=&\epsilon_{xx}+\epsilon_{yy}\;.
\end{eqnarray}
It then follows that the $\sigma_\alpha$ are given by
\begin{eqnarray}
\sigma_1&=&(\sigma_{xx}-\sigma_{yy})/2\;,\\
\sigma_2&=&(\sigma_{xy}+\sigma_{yx})/2\;,\\
\sigma_3&=&(\sigma_{xx}+\sigma_{yy})/2.
\end{eqnarray}
 In this study, we shear the system in mode $2$ so we are mainly interested in its corresponding quantities.
So we have:  $\epsilon_{xx}=\epsilon_{yy}=0$ and $\epsilon_{xy}=\epsilon_{yx}=\frac{1}{2}$ which gives $\epsilon_{2}=\epsilon_{xy}+\epsilon_{yx}=1$ and, following equation~\ref{eq:eshelby_tensor_circle-lameS}, $\gamma_2=\gamma_{xy}+\gamma_{yx} = \frac{\lambda+3\mu}{2\lambda+4\mu}$.
The stress in the circular inclusion, the constrained stress, $\sigma_{c}$, is proportional only to the elastic part of the strain, so we have $\sigma_{c}=\mu(\gamma_2-\epsilon_2)=\mu(\gamma_2-1)$ which gives:
\begin{equation}
    \sigma_{c} = -\frac{\mu}{2}\frac{\lambda+\mu}{\lambda+2\mu}=-\mu\frac{\alpha}{4}\;,
\label{eq:sigmac_continuum}
\end{equation}
% Where $\mu$ is the shear modulus, and $\lambda$ is the other Lam\'e parameter.
where we have defined the dimensionless combination of Lam\'e parameters: 
\begin{equation}
    \alpha = 2\frac{\lambda+\mu}{\lambda+2\mu}=2\frac{K}{K+\mu}\;.
    \label{eq:alpha}
\end{equation}
Note that the negative sign in equation~\ref{eq:sigmac_continuum} indicates that the inclusion experiences a shear stress in the opposite direction to the positive plastic strain $\epsilon_{2}$ that was prescribed. 
The amplitude of this constrained stress is controlled by the dimensionless parameter  $\alpha$, which, at fixed $\mu$, monotonically increases with the compression modulus, $K$, and saturates at $\alpha=2$ for incompressible materials.

It is important to note that, by symmetry, we would obtain precisely the same constrained stress for any orientation of the shear: be it in mode-1 with $\epsilon_1=1$, mode-2 with $\epsilon_2=1$, or any combination of them.  
Furthermore, if we impose an eigenstrain with pure shear and no dilatancy, so that $\epsilon_3=0$, then there will be no resulting dilatant strain, so $\gamma_3$ will be zero. 
Finally, and most importantly, in the \emph{continuum} solution presented here for \emph{an infinte medium without a boundary}, the entire solution both inside the inclusion and in the matrix for mode-2 will be precisely equal to the solution for mode-1 rotated by $45$ degrees and vice versa.
The solution in a periodic square cell discretized with finite differences on a square grid presented below will break this symmetry in subtle ways and mode-1 and mode-2 will be closely related to each other but not precisely equivalent as they are in the infinite continuum.

\subsection{Staggered grid discretization}

In our discretization of the Eshelby problem, a 2D material is divided into square tiles, and the total energy, $U$, is defined as the sum of the local energies of the tiles.
That is: $U=\sum_{IJ}U[I,J]$, where $I$ and $J$ are the spatial indices of the tiles with $0\leq I<L$ and $0\leq J<L$.
The total strain at a tile, $\gamma[I,J]$, is additively decomposed into an elastic part, $\epsilon_e[I,J]$, and a plastic part, $\epsilon[I,J]$, 
where the plastic part is to be identified with the eigenstrain in the continuum Eshelby problem.
So we have:
\begin{equation}
\gamma_1[I,J]=\epsilon_{\text{e}1}[I,J]+\epsilon_1[I,J]\;.
\label{eq:gamma_two_parts}
\end{equation}
We define the local energy using only the elastic strain:
\begin{equation}
\begin{split}
 U[I,J]&=\frac{\mu}{2} ( (\gamma_1[I,J]-\epsilon_{1}[I,J])^2\\
        &+(\gamma_2[I,J]-\epsilon_{2}[I,J])^2)+\frac{K}{2}(\gamma_3[I,J])^2   \;.
\end{split}
\label{eq:energy_eshelby}
\end{equation}
So then the shear stress for modes $1$ and $2$, $\sigma[I,J]$, is:
\begin{equation}
\sigma[I,J]=\mu\epsilon_e[I,J]=\mu(\gamma[I,J]-\epsilon[I,J]) \;.
\label{eq:sigma_from_elastic}
\end{equation}
The discretized Eshelby problem is then to find the total strain, $\gamma$, and the displacement, $u$, from which it was derived, that minimize $U$ for a given $\epsilon$.
%subject to the kinematic compatibility condition -- the requirement that the strain be the derivative of some displacement field -- and a given $\epsilon$. 
%To ensure that $\gamma$ satisfies kinematic compatibility we explicitly minimize $U$ with respect to the displacement field, $u$.
To obtain $\gamma$ from $u$, we use the so-called staggered difference scheme ~\cite{10.1785/BSSA0660030639,Levander:1988ue,Randall1991MultipoleBA,Virieux:1986tb,10.1785/BSSA0860041091,Schneider:2016tn}: 
\begin{equation}
\begin{split}
\gamma_{xx}[I,J]&=u_x[I+1,J]-u_x[I,J]\;, \\  
\gamma_{xy}[I,J]&=u_y[I,J]-u_y[I-1,J]\;, \\
\gamma_{yx}[I,J]&=u_x[I,J]-u_x[I,J-1]\;,  \\  
\gamma_{yy}[I,J]&=u_y[I,J+1]-u_y[I,J]\;.
\end{split}
\label{eq:stag}
\end{equation}

We put \ref{eq:stag} into \ref{eq:energy_eshelby} and solve for the $u$  that minimizes the energy and then substitute the solution, $u$, into \ref{eq:stag} to get the strain from which we can find the stress by using ~\ref{eq:sigma_from_elastic}.
This is completely equivalent to minimizing the total energy with respect to the total strain field \emph{subject to the constraint that the total strain field be kinematically compatible -- i.e. be the derivative of a displacement field}.
The staggered grid scheme implicitly defines the notion of kinematic compatibility.
% Because of the linearity of the problem, we can find the solution, $\gamma$ to a prescribed $\epsilon_p$ by convolving it with the solution, $G$, to a discretized delta function $\epsilon_p[I,J]=\delta_{I0}\delta_{J0}$, then we have:
% \begin{equation}
% \begin{split}
% \gamma_{1}[I,J]&=\sum_{IJ}G_{11}[M-I,N-J]\epsilon_{p1}[M,N]\\ 
%                &+G_{12}[M-I,N-J]\epsilon_{p2}[M,N]
% \end{split}
% \end{equation}
% \begin{equation}
% \begin{split}
% \gamma_{2}[I,J]&=\sum_{IJ} G_{21}[M-I,N-J]\epsilon_{p1}[M,N]\\ 
%                &+G_{22}[M-I,N-J]\epsilon_{p2}[M,N]
% \end{split}
% \end{equation}
To find the so-called Eshelby kernel, $G[I,J]$, we consider the special case where the prescribed $\epsilon[I,J]$ is the discrete delta function: $\epsilon[I,J]=\delta_{I0}\delta_{J0}$. 
Let $G_{\alpha\beta}$ be the $\alpha$-th Voigt strain resulting from an imposed  eigenstrain in mode $\beta$.
So then $G_{11}$ is the $\gamma_1$ strain resulting from an imposed $\epsilon_1$ eigenstrain,  $G_{12}$ is the $\gamma_1$ strain resulting from an imposed $\epsilon_2$ eigenstrain, $G_{21}$ is the $\gamma_2$ strain resulting from an imposed $\epsilon_1$ eigenstrain, and $G_{22}$ is the $\gamma_2$ strain resulting from an imposed $\epsilon_2$.
We use a staggered grid scheme here as using a second order centered difference scheme gives pathological behavior for $G_{12}$ and $G_{21}$, however, the results for $G_{11}$ and $G_{22}$ are equivalent to what one would obtain in a second order centered difference scheme. 
To solve for $G$, it is easiest to work in Fourier space.

We use the following Fourier transform conventions~\cite{press2007numerical}.
\begin{equation}
f[I,J]=\frac{1}{L^2}\sum_{m,n=0}^{L-1}\tilde{f}[m,n]\exp\left[-i2\pi (Im+Jn)/L \right]
\label{eq:fft}
\end{equation}
where:
\begin{equation}
\tilde{f}[m,n]=\sum_{I,J=0}^{L-1}f[I,J]\exp\left[+i2\pi (Im+Jn)/L \right]
\label{eq:ifft}
\end{equation}
After differentiating the total energy with respect to the displacement field and performing some algebra to solve for the energy minimum, we obtain $\tilde G$:
%  \begin{eqnarray}
% \tilde{G}_{11}[m,n]&=&-\frac{2(2\lambda+3\mu)\Delta_x^2\Delta_y^2+\mu(\Delta_x^4+\Delta_y^4)}{(\lambda+2\mu)\mathcal{D}}\\
% \tilde{G}_{21}[m,n]&=&\frac{2(\lambda+\mu) q_{x-} q_{y-} (\Delta^2_x-\Delta^2_y)}{(\lambda+2\mu)\mathcal{D}}\\
% \tilde{G}_{12}[m,n]&=&\frac{2(\lambda+\mu) q_{x+} q_{y+} (\Delta^2_x-\Delta^2_y)}{(\lambda+2\mu)\mathcal{D}}\\
% \tilde{G}_{22}[m,n]&=&\frac{2\lambda\Delta^2_x\Delta^2_y-(\lambda+2\mu)(\Delta^4_x+\Delta^4_y)}{(\lambda+2\mu)\mathcal{D}}
% \end{eqnarray}
 \begin{eqnarray}
\tilde{G}_{11}[m,n]&=&\frac{2\lambda+3\mu}{\lambda+2\mu}\tilde A[m,n]+\frac{\mu}{{\lambda+2\mu}}\tilde B[m,n] \label{eq:G11tilde_original}\\
\tilde{G}_{21}[m,n]&=&-\frac{2(\lambda+\mu) q_{x-} q_{y-} (\Delta^2_x-\Delta^2_y)}{(\lambda+2\mu)\mathcal{D}}\label{eq:G21tilde_original}\\
\tilde{G}_{12}[m,n]&=&-\frac{2(\lambda+\mu) q_{x+} q_{y+} (\Delta^2_x-\Delta^2_y)}{(\lambda+2\mu)\mathcal{D}}\label{eq:G12tilde_original}\\
\tilde{G}_{22}[m,n]&=&-\frac{\lambda}{\lambda+2\mu}\tilde A[m,n]+\tilde B[m,n] \label{eq:G22tilde_original}
\end{eqnarray}
for $(m,n)\neq (0,0)$ and $\tilde{G}[0,0]=0$.
Setting $\tilde{G}[0,0]=0$ simply fixes the spatial average of the strain to be $0$.
A non-zero average strain would shift the total energy but not change the displacement field which minimizes it.
Furthermore, any strain field defined by the staggered grid procedure in equations~\ref{eq:stag} must have zero spatial average if the boundaries of the periodic cell are assumed to be held fixed, and this condition is not implied by the algebra in Fourier space and must be specified separately.  
Here: $(m,n)$ are Fourier indices; 
\begin{eqnarray}
\Delta^2_x&=&q_{x+}q_{x-}\label{eq:Deltax^2}\\
\Delta^2_y&=&q_{y+}q_{y-}\label{eq:Deltay^2}
\end{eqnarray}
are the Fourier transforms of the second ordered centered difference operators in the $x$ and $y$ direction; 
\begin{equation}
\mathcal{D}=(\Delta_x^2+\Delta_y^2)^2
\label{eq:curlyD}
\end{equation}
is the Fourier transform of the bi-Laplacian.
We have defined:
\begin{equation}
\tilde A[m,n]=\frac{2\Delta_x^2\Delta_y^2}{\mathcal{D}}
\label{eq:Atilde}
\end{equation}
\begin{equation}
\tilde B[m,n]=\frac{\Delta_x^4+\Delta_y^4}{\mathcal{D}}
\label{eq:Btilde}
\end{equation}
for $(m,n)\neq (0,0)$ and $\tilde{A}=0, \tilde{B}=0$ for $(m,n)=(0,0)$ so that $\tilde{G}[0,0]=0$ to have zero average total strain in real space.
The $q_{x\pm}$ and $q_{y\pm}$ are the Fourier transforms of the forward and backward difference operators in the $x$ and $y$ directions~\cite{Schneider:2016tn}
\begin{eqnarray}
q_{x+}[m,n]&=&-i (+\exp[+2\pi i m/L]-1)\\
q_{x-}[m,n]&=&-i (-\exp[-2\pi i m/L]+1)\\
q_{y+}[m,n]&=&-i (+\exp[+2\pi i n/L]-1)\\
q_{y-}[m,n]&=&-i (-\exp[-2\pi i n/L]+1)
\end{eqnarray}

In the continuum Eshelby problem of a circular inclusion in an infinite linear elastic medium, there is a high degree of symmetry which is only approximate in the present model where the symmetry is broken both by the square periodic cell and the discrete square tiling.
In the infinite continuum result, the $G_{11}$ and $G_{22}$ fields can be obtained from each other simply by a $45$-degree rotation.
In the equivalent continuum calculation, one would have $\tilde{A}=2\Delta_x^2\Delta_y^2/\mathcal{D}\rightarrow 2\cos^2(\theta)\sin^2(\theta)$ and $\tilde{B}=(\Delta_x^4+\Delta_y^4)/\mathcal{D}\rightarrow \cos^4(\theta)+\sin^4(\theta)$ where $\theta$ is the angle of the wavevector in the $x,y$ plane.
One would then have: $\tilde{A}+\tilde{B}=2\cos^2(\theta)\sin^2(\theta)+\cos^4(\theta)+\sin^4(\theta)=1$.
In the present discrete square-cell case, $\tilde A[m,n]$ and $\tilde B[m,n]$ are related by: 
\begin{equation}
  \tilde A[m,n]+\tilde B[m,n]= 1-\delta_{m0}\delta_{n0}
  \label{eq:Atilde+Btilde}
\end{equation}
In other words, $\tilde{A}$ and $\tilde{B}$ sum to $1$ at every point in Fourier space except the origin.
So in real space
 \begin{equation}
  A[I,J]+ B[I,J]=\delta_{I0}\delta_{J0}-\frac{1}{L^2}
  \label{eq:A+B}
\end{equation}

Substituting ~\ref{eq:Atilde+Btilde} into ~\ref{eq:G11tilde_original} and ~\ref{eq:G22tilde_original} and recalling the definition of $\alpha$ from equation ~\ref{eq:alpha}, we get
% \begin{eqnarray}
% \tilde{G}_{11}[m,n]&=&\frac{2\lambda+2\mu}{\lambda+2\mu}\tilde A[m,n]+\frac{\mu(1-\delta_{m0}\delta_{n0})}{\lambda+2\mu}\\
% \tilde{G}_{22}[m,n]&=&-\frac{2\lambda+2\mu}{\lambda+2\mu}\tilde A[m,n] + (1-\delta_{m0}\delta_{n0})
% \end{eqnarray}

\begin{eqnarray}
\tilde{G}_{11}[m,n]&=&\alpha\tilde A[m,n]+\frac{\mu(1-\delta_{m0}\delta_{n0})}{\lambda+2\mu}\label{eq:G11tilde_of_A_alpha}\\
\tilde{G}_{22}[m,n]&=&-\alpha\tilde A[m,n] + (1-\delta_{m0}\delta_{n0})\label{eq:G22tilde_of_A_alpha}
\end{eqnarray}

in Fourier space, and in real space
 \begin{eqnarray}
{G}_{11}[I,J]&=&\alpha A[I,J]+\frac{\mu(\delta_{I0}\delta_{J0}-\frac{1}{L^2})}{(\lambda+2\mu)}\label{eq:G11_of_A_alpha}\\
{G}_{22}[I,J]&=&-\alpha A[I,J] + \delta_{I0}\delta_{J0}-\frac{1}{L^2}\label{eq:G22_of_A_alpha}
\label{eq:g22Real}
\end{eqnarray}
This implies that as $L\to\infty$, other than precisely at the origin, 
$G_{11}\approx - G_{22}\approx\alpha A[I,J]$.
$G_{11}$ is precisely equal to $-G_{22}$ in the continuum solution, whereas here it is only approximate and involves a correction term of order $1/L^2$ which is related to the boundary conditions in the discrete finite system that $G$ have zero spatial average.
% Which means that, except at the origin, as $\lambda \to \infty$ in the incompressible limit, $G_{11}\approx$ $-G_{22}\approx 2 A[I,J]$ in the limit of $L\to\infty$.
% The solution in real space is then given by taking a Fourier transform of (5):
% \begin{equation}
%     \epsilon_{\alpha,\beta}[I,J]=\frac{\epsilon_{p\mu\nu}}{L}\sum_{m=0}^{L-1}\sum_{n=0}^{L-1} \text{exp} \left[ -i2\pi \frac{mI+nJ}{L}\right]\tilde G_{\alpha\beta\mu\nu}[m,n]
% \end{equation}

%%%%%%%%%%%%%%%%%%%%%%%%%%%%%%%%%%%%%%%%%%%%%%%%
%%%%%%%%%%%%%%%%%%%%%%%%%%%%%%%%%%%%%%%%%%%%%%%%
\begin{figure}[h!]
\includegraphics[width = 0.45\columnwidth]{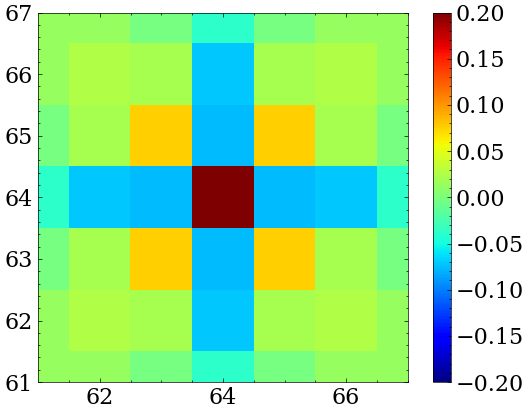}%
\includegraphics[width = .45\columnwidth]{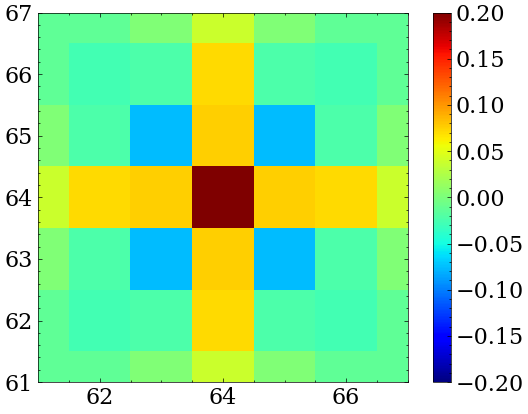}%
\caption{The real space $G_{11}$ (left) and  $G_{22}$ (right) near the origin for a system of size $L=128$ and $\lambda/\mu=4$.
Other than at the origin, $G_{11}=-G_{22}$ apart from a correction term which vanishes as $1/L^2$.
}
\label{fig:realSpaceG22}
\end{figure}
%%%%%%%%%%%%%%%%%%%%%%%%%%%%%%%%%%%%%%%%%%%%%%%%
%%%%%%%%%%%%%%%%%%%%%%%%%%%%%%%%%%%%%%%%%%%%%%%%
In figure \ref{fig:realSpaceG22}, we show $G_{11}$ and $G_{22}$ in real space near the origin for $\lambda/\mu=4$ for a system of size $L=128$.
Note that they are approximately equal in magnitude but opposite in sign other than at the origin, as in the continuum case.
However, the square lattice breaks the invariance under rotation by $\pi/4$ enjoyed by the continuum solution.

%%%%%%%%%%%%%%%%%%%%%%%%%%%%%%%%%%%%%%%%%%%%%%%%
%%%%%%%%%%%%%%%%%%%%%%%%%%%%%%%%%%%%%%%%%%%%%%%%
% \begin{figure}[h!]
% \includegraphics[width = 1.0\columnwidth]{G22_of_x.png}%
% \caption{Eshelby kernel $G_{22}(x,y=0)$ evaluated along the x axis for the first five adjacent sites for $\lambda=4,8,20,200$.}
% \label{fig:G22_of_x}
% \end{figure}
%%%%%%%%%%%%%%%%%%%%%%%%%%%%%%%%%%%%%%%%%%%%%%%%
%%%%%%%%%%%%%%%%%%%%%%%%%%%%%%%%%%%%%%%%%%%%%%%%

%%%%%%%%%%%%%%%%%%%%%%%%%%%%%%%%%%%%%%%%%%%%%%%%
%%%%%%%%%%%%%%%%%%%%%%%%%%%%%%%%%%%%%%%%%%%%%%%%
\begin{figure}[h!]
\includegraphics[width = 1.0\columnwidth]{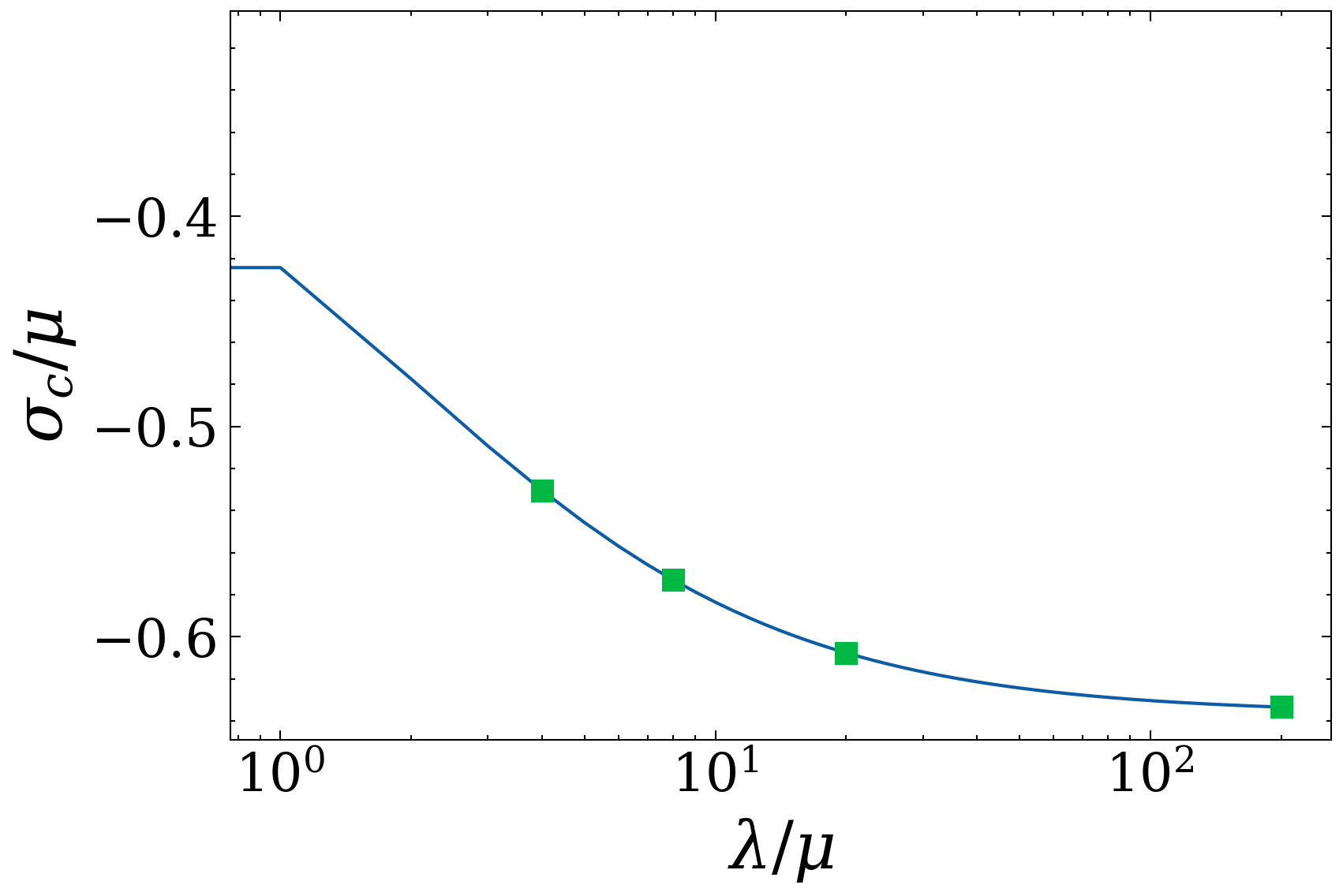}%
\caption{The constrained stress $\sigma_c$ in mode-2 vs. $\lambda/\mu$. 
The squares are the discretized solution for a system size $L=128$.
The solid line is the continuum solution (multiplied by $4A[0,0]=1.2732$) for a circular inclusion in an infinite medium.}
\label{fig:backstress}
\end{figure}
%%%%%%%%%%%%%%%%%%%%%%%%%%%%%%%%%%%%%%%%%%%%%%%%
%%%%%%%%%%%%%%%%%%%%%%%%%%%%%%%%%%%%%%%%%%%%%%%%

% In figure \ref{fig:realSpaceG22}, we show $G_{22}$ in real space near the origin for $\lambda=4$ (left) and $\lambda=200$ (right). 
% Going from the low value of $\lambda=4$ on the left to $\lambda=200$ we notice that the stress intensity increase in both senses in the vicinity of the transformed site and it dies quickly (around the fifth site) as shown in figure \ref{fig:G22_of_x}.

The constrained stress, $\sigma_c$, in the Eshelby problem is obtained by subtracting off the plastic piece, $\epsilon[0,0]=1$, from the total strain at the origin and multiplying by $\mu$. 
% So, for mode-1:
% \begin{eqnarray}
% \sigma_c&=&\mu(G_{11}[0,0]-1)\\
%   &=&\mu\left(\frac{\lambda+\mu}{2(\lambda+2\mu)}(2A[0,0]-1)-\frac{\mu}{L^2(\lambda+2\mu)}\right) \\
%   &=&\mu(G_{11}[0,0]-1)\\
%   &=&\mu\left(\frac{\alpha}{4}(2A[0,0]-1)-\frac{\mu}{L^2(\lambda+2\mu)}\right)
% \end{eqnarray}
So for mode-2:
\begin{eqnarray}
\sigma_c&=&\mu(G_{22}[0,0]-1)\\
  &=&-\mu\left(\alpha A_0+\frac{1}{L^2}\right) 
  \label{eq:sigmac2}
\end{eqnarray}
where
\begin{equation}
    A_0 = A[0,0]
    \label{eq:A0}
\end{equation}
%Comparing $\sigma_c$ in the discretized to the continuum solution it is easy to see that, if we neglect the $\frac{1}{L^2}$ term which vanishes as $L\to\infty$, the ratio between the discretized and the continuum versions of $\sigma_c$ is simply $(2A[0,0]-1)$ for mode-1, and $4A[0,0]$ for mode-2, so the discretized and the continuum $\sigma_c$ both have the same dependence on $\lambda$ and $\mu$.
% If we neglect the $\frac{1}{L^2}$ terms, we see that $\sigma_c$, in both mode-1 and mode-2, has the same dependence on the Lame parameters as in the continuum case.
% In mode-1 $\sigma_c$ is simply $(2A_0-1)$ times the continuum result, while for mode-2, $\sigma_c$ is $4A_0$ times the continuum result. 
(A similar calculation for  
mode-$1$ gives: $\sigma_c=-\mu\alpha(1-2A_0)/2+\mathcal{O}(1/L^2)$.)
If we neglect the $1/{L^2}$ terms, we see that $\sigma_c$ has the same dependence on the Lame parameters as in the continuum case.
In mode-2, $\sigma_c$ is $4A_0\approx 1.2732$ times the continuum-disk result, while in mode-1, $\sigma_c$ is $4(1/2-A_0)\approx 1.4536$ times the continuum-disk result. 
In figure \ref{fig:backstress}, we plot the discretized $\sigma_c$ in mode-2 (equation~\ref{eq:sigmac2}) for $L=128$ and the continuum $\sigma_c$ for a circular inclusion (equation~\ref{eq:sigmac_continuum}) multiplied by a prefactor of $4A_0$.
For convenience,  we have defined the particular dimensionless ratio of elastic moduli which emerges from the continuum Eshelby calculation as $\alpha$; it is the only combination of elastic moduli that will enter into our elasto-plastic model specified below.
$\alpha$ grows monotonically from zero at $K/\mu=0$ (a maximally negative Poisson ratio, completely compressible material) to $2$ at $\mu/K=0$ (an incompressible material). 

% For $L=128$, $A[0,0]=0.317674$, so the total strain is given by $G_{22}[0,0]\approx-\frac{2\lambda+2\mu}{\lambda+2\mu}(0.317674)+1$ and the constrained strain $\sigma_c\approx-\frac{2\lambda+2\mu}{\lambda+2\mu}(0.317674)$ where we ignore the negligible value $1/L^2$.
% Therefore, $G_{22}[0,0]$ ($\sigma_c$) is higher (lower) for lower values of $\lambda$, which means that the surrounding material is less constraining (exerts less opposing stress) to the transformation at the inclusion  for lower values of $\lambda$.

% So for two different materials with the same $\mu$ but different $\lambda$, an Eshelby inclusion with a given eigenstrain will generate shear stresses of higher intensity in the material with larger $\lambda$ — both inside the inclusion and in the matrix.  This is true  both for a circular inclusion in  the continuum mechanics calculation or a single site inclusion in the staggered grid framework.
% $\lambda$ bigger implies $\mathbf{S}$ bigger.
% $\mathbf{S}$ bigger implies $\sigma$ bigger.
% $\sigma$ bigger implies surface tractions bigger

$\sigma_c$ is negative which reflects the fact that, in the canonical Eshelby problem where the matrix and inclusion are both initially stress-free before the transformation -- after the transformation, the inclusion feels a stress from the matrix pushing it backwards away from its new stress-free configuration at $\gamma=1$ where the stress would be zero.
Since $\sigma_c$ is negative and a decreasing function of $\lambda/\mu$, its \emph{magnitude} increases.
This indicates larger surface tractions applied by the matrix to constrain the inclusion.
As we will see below, these larger backward stresses in the Eshelby problem lead to macroscopic yielding behavior at lower values of applied strain and stress despite the fact that the elementary yield strain of the local elements is held fixed in the model.

\section{Model}
\label{sec:model}

%%%%%%%%%%%%%%%%%%%%%%%%%%%%%%%%%%%%%%%%%%%%%%%%
%%%%%%%%%%%%%%%%%%%%%%%%%%%%%%%%%%%%%%%%%%%%%%%%
% \begin{figure}[h!]
% \includegraphics[width = 1.0\columnwidth]{parabola3.png}%
% \caption{Local energy and at any tile, $U$, vs. the total strain at the tile, $\gamma$.
% Suppose a tile is initially in the middle parabola (red dot) with $\gamma=1$, $\epsilon=0$, and $\sigma=\gamma-\epsilon=1$, before yielding. 
% Because $\gamma$ is at threshold, the tile will then yield and jump to the next parabola on the right (green dot) and $\epsilon$ is incremented by $2$.
% The new $\gamma$ is determined by the Eshelby solution and it depends on $\lambda$.
% The new stress after yielding, $\sigma_0$, is what we call the back stress.}. 
% \label{fig:backstress_in_parabola}
% \end{figure}
%%%%%%%%%%%%%%%%%%%%%%%%%%%%%%%%%%%%%%%%%%%%%%%%
%%%%%%%%%%%%%%%%%%%%%%%%%%%%%%%%%%%%%%%%%%%%%%%%

%%%%%%%%%%%%%%%%%%%%%%%%%%%%%%%%%%%%%%%%%%%%%%%%
%%%%%%%%%%%%%%%%%%%%%%%%%%%%%%%%%%%%%%%%%%%%%%%%
\begin{figure}[h!]
\includegraphics[width = 1.0\columnwidth]{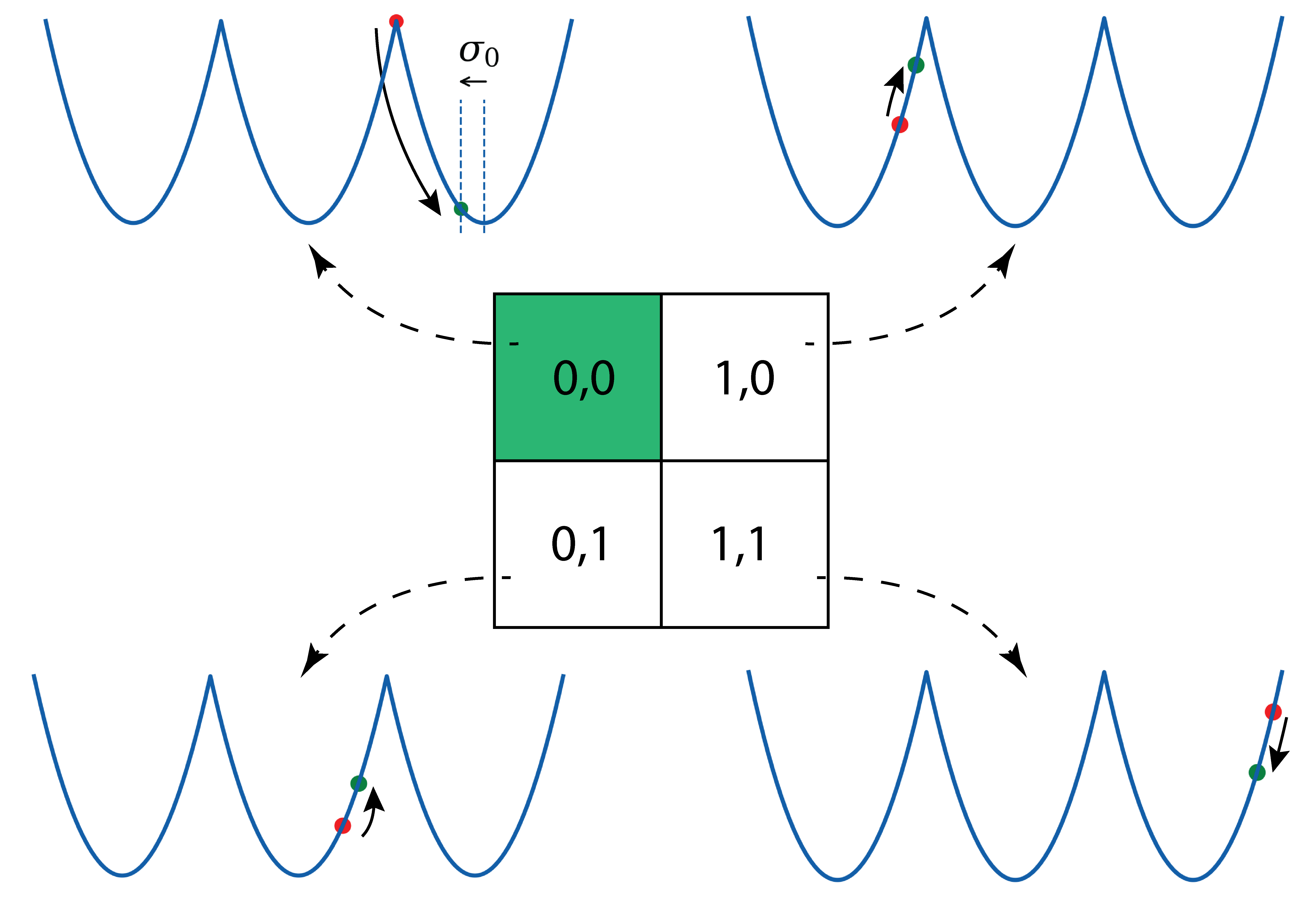}%
\caption{
A schematic of a single elementary shear transformation in the automaton at a site $[0,0]$, and the resulting changes in local strain and energy in a few nearby sites on the grid after re-equilibration.
The parabolas show the local energy, $U$, of the sites as a function of the total strain, $\gamma$.
%There is a transition at ach parabola ranges from $\gamma=2*n-1$ to $\gamma=2*n+1$ with the transition from one parabola to the next occurring at $\gamma=2n\pm1$.
The center of the $n$-th parabola is at $\gamma=2n$, and the transition between basin $n$ and basin $n\pm 1$ occurs at $\gamma=2n\pm 1$.
%The index, $n$, of the parabola a site is in is given by $\epsilon/2$ where $\epsilon$ is the plastic strain at the site.
%The stress at a site, $\sigma$, is just the signed distance from the current $\gamma$ at the site to $\gamma$ at the center of the current parabola (where $U$ is minimum).
The stress, $\sigma$, as a function of total strain, $\gamma$, is then $\sigma=\gamma-2F[\frac{\gamma}{2}+\frac{1}{2}]$ where $F$ is the floor function defined in the text.
The four different identical landscapes correspond to the four different tiles shown.
The red and green dots indicate local strain values before and after a shear transformation at the site labeled $[0,0]$.
After the transformation, sites $[1,0]$ and $[0,1]$ are at a higher energy and stress, site $[1,1]$ is at a lower energy and stress, while the transforming site, site $[0,0]$, is at a much lower -- but non-zero -- energy and at a \emph{negative} stress.
}. 
\label{fig:backstress_in_parabola2}
\end{figure}
%%%%%%%%%%%%%%%%%%%%%%%%%%%%%%%%%%%%%%%%%%%%%%%%
%%%%%%%%%%%%%%%%%%%%%%%%%%%%%%%%%%%%%%%%%%%%%%%%
To implement an elasto-plastic model, we suppose that any site in the material has the same piecewise quadratic strain-energy function as given by equation~\ref{eq:energy_eshelby}, but where the plastic shear strain, $\epsilon_{1(2)}$, is determined self-consistently to connect the parabolas.
%To specify an evolution rule of the plastic strain, $\epsilon$, we define it as:
\begin{equation}
    \epsilon_{1(2)}=2F\left[\frac{\gamma_{1(2)}}{2}+\frac{1}{2}\right]
    \label{eq:floor}
\end{equation}
where $F[x]$ is the floor function that returns the greatest integer less than or equal to $x$.
This construction gives parabolas which each range from $\gamma=2n-1$ to $\gamma=2n+1$ with transitions from one parabola to the next occurring at $\gamma=2n\pm 1$ (figure \ref{fig:backstress_in_parabola2}).
Since the minima are separated by a strain of $2$, the plastic strain increment in the model must be $2$ as reflected in the prefactor of $2$ on the floor function.

This sets the microscopic condition for yielding (a transition from one parabola to the next) as $\gamma-\epsilon=\pm1$, and, implicitly this sets the microscopic yield strain to unity.
All quantities with units of strain emerging from the model should be understood to be measured in units of this microscopic yield strain.
We also set $\mu=1$ and measure all stresses and $\lambda$ and $K$ in units of $\mu$.
This piecewise-quadratic strain energy function then gives a single-valued Hamiltonian for the system which depends on the displacement field implicitly via the total strain and is piece-wise quadratic in the displacement field.
The discretized Eshelby solution then provides a way to find and enumerate energy minima of this Hamiltonian.
To use the Hamiltonian to define an evolution protocol for an integer automaton we proceed as follows:
%The model evolution protocol is then: 
i) For a given $\gamma$ and $\epsilon$, increment (or decrement) $\epsilon$ by $2$ in all the sites with $|\gamma-\epsilon|>1$ and, simultaneously, update $\gamma$ everywhere in accordance with the discretized Eshelby solution.
ii) Repeat (i) until all sites are below threshold with $|\gamma-\epsilon|<1$.
At this point, the $\gamma$ field will be the derivative of a displacement field which minimizes the total Hamiltonian, and, as such will automatically be compatible. 
iii) Advance the imposed total strain homogeneously until one site is precisely at threshold with $|\gamma-\epsilon|=1$.
iv) Repeat.
%iv) Repeat steps (i)-(iii).
%Again, we set $\mu=1$ in this study so $\sigma$ and $\epsilon_e$ are numerically equal.
We initialize the systems by adding a random plastic field of $\pm 2$ at each site (in both mode1 and mode 2) and running the automaton evolution.
We have checked that repeating this initialization process again on a previously initialized configuration does not affect our results qualitatively, and our initial state should be thought of as a rapidly quenched glass from a high temperature liquid with a high degree of disorder.

The stress increment at site $[I,J]$ due to a transformation at site $[K,L]$ is given by:
\begin{equation}
\Delta\sigma[I,J]=-2\alpha A[K-I,L-J]-\frac{2}{L^2}
\label{eq:deltaSigma}
\end{equation}
This expression is valid for any site in the system \emph{including the transforming site itself}.
If we recall that $\sum_{IJ}A[I,J]=0$ by construction, this makes it clear that there is a quantum of stress relaxation associated with a single transformation in our model: $\Delta \sigma= -2/L^2$ which is independent of $\alpha$.
Here, $\sigma$ with no lattice indices refers to the spatial average: $\sigma=\frac{1}{L^2}\sum_{IJ}\sigma[I,J]$.
It is only the amplitude of stress redistribution, rather than the quantum of relaxation, which depends on $\alpha$, and it is, in fact, simply proportional to $\alpha$.
A word of caution is appropriate here: the $\Delta\sigma[I,J]$ defined in equation~\ref{eq:deltaSigma} excludes the $\delta_{IK}\delta_{JL}$ which appears in equation~\ref{eq:g22Real}, as we must subtract off the plastic strain arising in the Eshelby problem to obtain the elastic strain at the origin.

In figure \ref{fig:backstress_in_parabola2}, we show a schematic of a single elementary shear transformation at a site $[0,0]$.
In the figure, we show only the transforming site and three neighboring sites, but it is to be understood that the transformation will affect the strain values at all sites in the plane.
%Suppose that, initially, all sites are below threshold and we homogeneously advance $\gamma$ until only site [0,0] is at the forward threshold.
Suppose that, initially, site $[0,0]$ is at the threshold as we are loading the system in the positive direction.
%From the point of view of the Hamiltonian, the site would experience a discontinuous stress change from $+1$ to $-1$ as the local strain value crosses from $\gamma<1$ to $\gamma>1$. 
The total strains at the sites before the transition are  indicated by the red dots.
The location of all the red dots corresponds to an energy minimum of the Hamiltonian.
The red dots need not all be in the same basin; i.e. there may be plastic strain in the system, and local values of strain are determined by the values of the plastic strain throughout the entire system.   
%Because site [0,0] is at the positive threshold, it will transform by transitioning to the next parabola where $\epsilon$ is incremented by $2$.
As site $[0,0]$ crosses $\gamma=1$, the plastic strain, $\epsilon$, will be incremented by $2$.
After the event, the new total strains (and stresses) at all sites will be incremented/decremented according to the Eshelby solution \emph{including site $[0,0]$ itself!}
The updated strains are indicated by the green dots.
At site [0,0], $\gamma$ will be incremented and will be at the left half of the next parabola. 
The new stress at site [0,0] immediately after yielding, $\sigma_0$, is the so-called back stress, and it sits at a negative value when the site yields in the forward direction.
$\sigma_0$ is obtained from  $2\sigma_c$ by adding $1$ since $\sigma_c$ is the post-transformation stress at the transformed site in a system that is initially stress free:
\begin{equation}
  \sigma_0=2\sigma_c+1 = 1-2\alpha A_0-\frac{2}{L^2}
  \label{eq:sima0}
\end{equation}
%Because of the symmetry in the Eshelby solution, $\gamma$ at sites [1,0] and [0,1] will be incremented by the same magnitude and decremented at site [1,1], also by the same magnitude.
The locations of the red and green dots in the figure are determined from the exact values for $\lambda=200$ in a $128$-by-$128$ system. 
%but the other values of $\lambda$ we are considering in this study ($4,8$ and $20$) behave qualitatively in a similar way.

% and running step i) and ii) of the automaton evolution without advancing the imposed strain.
%We then cyclically shear the system in mode 2 (deformation modes are described in the appendix)
% Again, we define the back stress, $\sigma_0$, as the stress value of a site immediately after yielding at $\gamma-\epsilon=1$ (see figure \ref{fig:backstress_in_parabola}) so it is obtained from  $\sigma_c$ by adding $1$ since $\sigma_c$ is the shear stress in a transformed site that is initially stress free:
% \begin{equation}
%   \sigma_0=2\sigma_c+1
% \end{equation}

% where $\sigma_y$ is the value at which the site yields.
In our previous study \cite{elgailani2022anomalous}, we showed, in the incompressible limit, that $\sigma_0$ gives rise to a characteristic shape of the stress distribution and we argued that it can be used to predict the location of $\Gamma_0$.   
Here we show the same relation between $\sigma_0$ and $\Gamma_0$ follows for the finite compression modulus case, despite the dependence of $\sigma_0$ on $K/\mu$.
Furthermore, we show that a similar relation exists between $\sigma_0$ and $\sigma_y$, the steady state flow stress in steady-forward shear.
%In the following, we study the dependence of $\sigma_0$ on $\lambda$ and whether the predictions of $\Gamma_0$ based on $\sigma_0$ holds; it turns out, it holds with high accuracy.
%In the following, we study the dependence of $\sigma_0$ on $\lambda$.
%The relation between $\Gamma_0$ and $\sigma_0$ found in the incompressible study holds with high accuracy.

% \section{Results and discussion}
\section{Cyclic loading}
\label{sec:cyclic-loading}

In this section, first we present the loading curves for systems of various compressibility subjected to cyclic shear.
For cycling amplitudes below $\Gamma_0$, the terminal loading curves are trivial with no hysteresis.
For cycling amplitudes above $\Gamma_0$, we show that the hysteretic loading curves have a quiescent period devoid of any significant plastic activity followed by a rapid initiation of plastic activity after the strain exceeds a level, $\gamma_*$, and we explain the dependence of $\gamma_*$ on $\Gamma_0$. 
Next, we relate the rapid onset of plasticity at $\gamma_*$ to the form of the probability distribution of local stress values and explain how important features of that distribution are governed by the residual stress present at a local site after an Eshelby transformation.
Finally, we show how the compressibilty dependence of the residual stress in the Eshelby transformation perfectly accounts for the compressibility dependence of $\Gamma_0$.

\subsection{Strain and strain rate curves}
%%%%%%%%%%%%%%%%%%%%%%%%%%%%%%%%%%%%%%%%%%%%%%%%
%%%%%%%%%%%%%%%%%%%%%%%%%%%%%%%%%%%%%%%%%%%%%%%%
\begin{figure*}[h!]
\includegraphics[width = 2.\columnwidth]{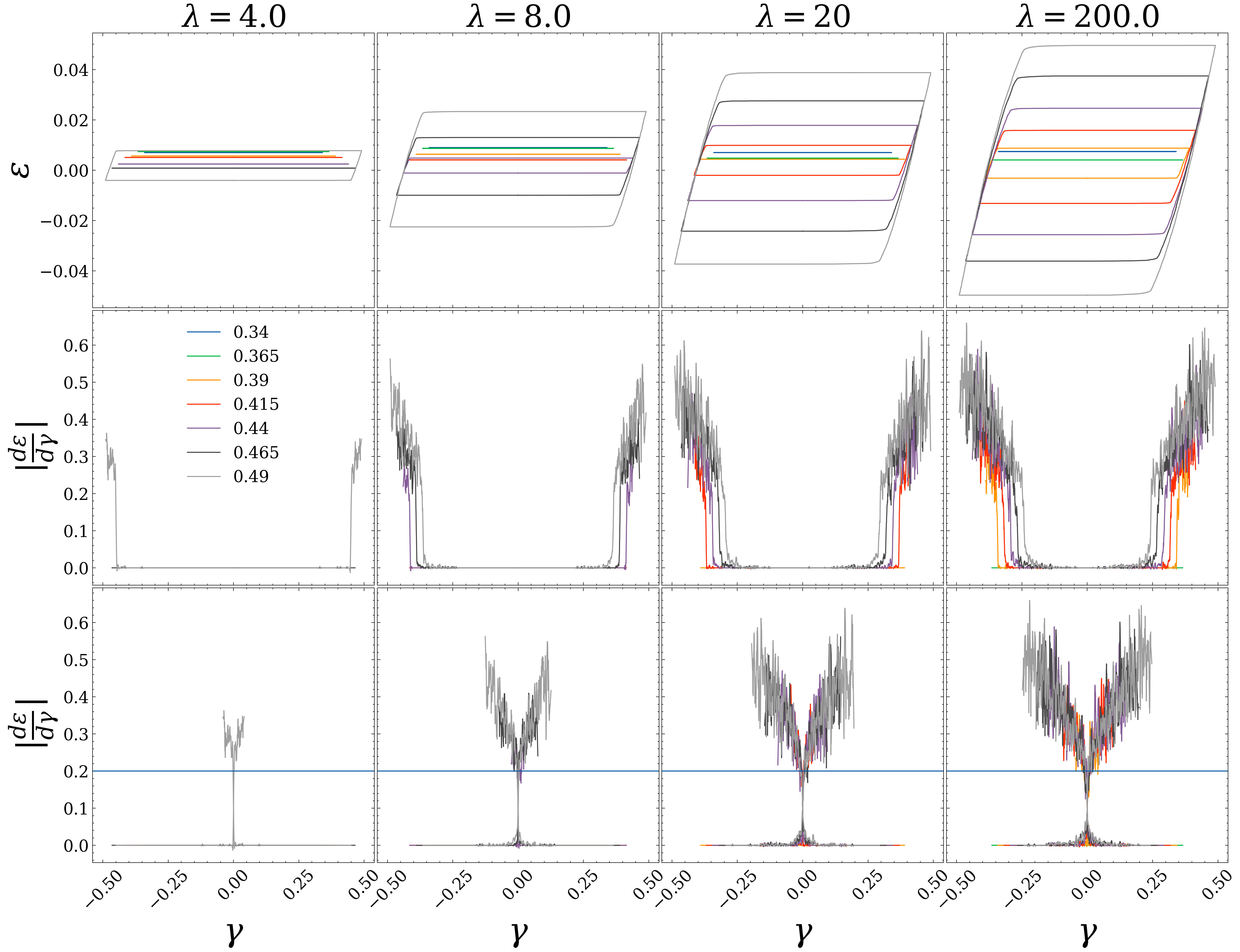}%
\caption{Top row: plastic strain $\epsilon$ (ensemble average) in the limit cycles vs. strain $\gamma$ for several different strain amplitudes $\Gamma$ and four different values of Lame parameter $\lambda$. 
Middle row: the strain rate $|\frac{d\epsilon}{d\gamma}|$ vs $\gamma$ for the curves in the top row. 
Bottom row: the strain rate $|\frac{d\epsilon}{d\gamma}|$ vs $\gamma-\gamma_*$ for the curves in the top row where $\gamma_*$ is the plasticity onset strain. The solid line in the bottom row is at $|\frac{d\epsilon}{d\gamma}|=0.2$  }
\label{fig:hysteresis}
\end{figure*}
%%%%%%%%%%%%%%%%%%%%%%%%%%%%%%%%%%%%%%%%%%%%%%%%
%%%%%%%%%%%%%%%%%%%%%%%%%%%%%%%%%%%%%%%%%%%%%%%%

In figure \ref{fig:hysteresis} (top), we show the ensemble average of the plastic strain, $\epsilon$, in the terminal limit cycles--after the transient has died away-- vs. the applied strain, $\gamma$, for several different cycling amplitudes, $\Gamma$, both below and above the transition value, $\Gamma_0$, for different values of $\lambda$:4, 8, 20, and 200.

We immediately notice the strong dependence of the hysteresis on the compressibility. 
Systems with larger $\lambda$ (smaller compressibility), have larger hysteresis amplitude.
For all $\lambda$ values, for $\Gamma<\Gamma_0$, the system is in the ER, where after the transient, $\epsilon$ is a flat line with no plastic events or energy dissipation. 
The exact location of the residual $\epsilon$, where the flat lines are located, is a non-trivial function of the cycling amplitude, $\Gamma$, but they all sit at a positive value. 
This symmetry breaking is a result of our arbitrary choice to load the virgin state in the forward direction first.
For the smallest $\lambda=4$ in our set, only the highest amplitude, $\Gamma=0.49$, is in the RPR as indicated by the hysteretic $\epsilon$ vs. $\gamma$ loop.
% For higher $\lambda$, lower amplitudes in the set are in the RPR [break down to few sentences].
As $\lambda$ increases, hysteresis starts at lower $\Gamma$.
For example, at $\lambda=8.0$, there are three amplitudes in the RPR: $0.44$, $0.465$, and $0.49$.
Similarly, for $\lambda=20$ and $200$ we have, respectively, $4$ and $5$ amplitudes in the RPR.
% This behavior indicates that systems with higher $\lambda$ yield at an earlier $\Gamma$.
We discuss the location, $\Gamma_0$, of the onset of hysteresis and its dependence on $\lambda$ in the text below.
% Furthermore, for a given $\Gamma$ in the RPR, systems with higher $\lambda$ are softer with a smaller onset strain ,$\gamma_*$, defined by setting a threshold at $|\frac{\epsilon}{\gamma}|>0.1$.

In figure \ref{fig:hysteresis} (middle), we plot the plastic strain rate, $|d\epsilon/d\gamma|$.
The strain rate curves all show a relatively sharp transition from zero to roughly $0.2$ at a characteristic value of strain, $\gamma_*$, which depends on both $\lambda$ and $\Gamma$.
We define $\gamma_*$ as the $\gamma$ value for which the strain rate first exceeds $0.1$, and we have checked that $\gamma_*$ is not very sensitive to this precise choice as the jump in $\frac{d\epsilon}{d\gamma}$ is rather sharp.
% We note that $\gamma_*$ depends on both and $\lambda$ and discuss this dependence further below.
In the bottom row, we shift the curves horizontally and show $|\frac{d\epsilon}{d\gamma}|$ vs. $\gamma-\gamma_*$.
After shifting, the curves can be seen to have roughly the same shape.
% {\bf XXX-AE: repreated sentence?} 
%There is a sharp jump from zero to approximately $0.2$ at the onset of plasticity, reminiscent of first order phase transition.
%The size of the jump is independent of $\lambda$ and $\Gamma$ and roughly equalt to $0.2$.
While the size of the jump is independent of $\lambda$ and $\Gamma$, the sharpness of the jump depends on $|\Gamma-\Gamma_0|$.
The larger the $\Gamma-\Gamma_0$, the more rounded the transition is, with more plasticity occurring at $|\gamma|<\gamma_*$.
This regime of small amounts of plasticity at $|\gamma|<\gamma_*$ is reminiscent of creep in other systems \cite{Merabia:2016um, Castellanos:2019ur, Liu:2021un, Ferrero:2021ta, Popovic:2022wt}.
It is interesting to note that the systems which are cycled at $\Gamma$ just slightly above $\Gamma_0$ have virtually no creep and no plasticity until a sudden burst of plasticity at the turning point $\gamma\approx\Gamma$ where the sense of the driving strain is reversed.
For example, the gray curve on the left at $\lambda=4, \Gamma=0.49$ has virtually no plasticity until the sudden burst near the turning point at $\gamma\approx 0.49$.

% In the RPR, if we ignore the creeping regime below $\gamma_*$ since it does not contribute much to the total plasticity, we find that  $|\frac{d\epsilon}{d\gamma}|$ start from zero by jumping to a finite value, approximately $0.2$, indicative of a first order transition, independent of $\lambda$ and $\Gamma$.
% The systems which are cycled at $\Gamma$ values closest to the $\Gamma_0$ threshold show the sharpest transition, while systems cycled at larger $\Gamma$ values show a more rounded transition with some noticeable plasticity below $\gamma_*$ reminiscent of creep seen in xxx\cite{}

\subsection{Plasticity onset at $\gamma_*$}

%%%%%%%%%%%%%%%%%%%%%%%%%%%%%%%%%%%%%%%%%%%%%%%%
%%%%%%%%%%%%%%%%%%%%%%%%%%%%%%%%%%%%%%%%%%%%%%%%
\begin{figure}[h!]
\includegraphics[width = 1.\columnwidth]{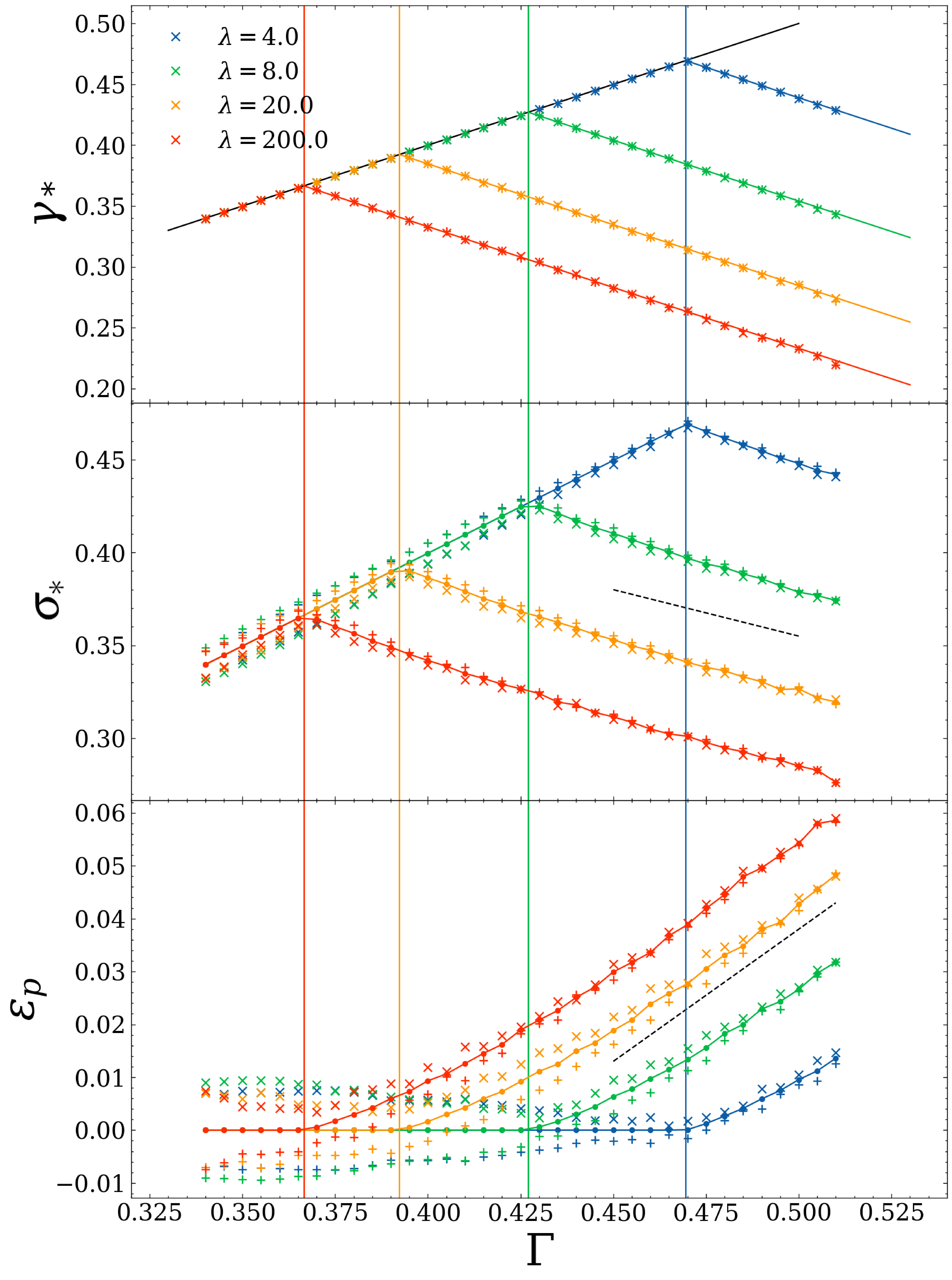}%
\caption{Top row: The (negative) onset strain, $\pm\gamma_*^{+(-)}$, in the forward (and reverse) direction for different values of the first Lame parameter $\lambda$. 
The solid lines are $\gamma_*=\Gamma_0-(\Gamma-\Gamma_0)$ with $\Gamma_0$ determined by taking the average $(\Gamma+\gamma_*)/2$ for a given $\lambda$.
Middle row: The (negative) stress at onset, $\pm\sigma_*^{+(-)}$, in the forward (and reverse) direction. 
The dashed line is a guide to eye with a slope = -1/2.
Bottom row: The (negative) plateau strain, $\pm\epsilon_p^{+(-)}$, in the forward and (reverse) direction.
The dashed line is a guide to eye with a slope = 1/2.
The vertical lines are at $\Gamma_0$ for the different $\lambda$ values. 
In all cases, the crosses and pluses are for the forward and reverse directions, respectively. The dots in b) and c) mark the averages.}
\label{fig:fig2}
\end{figure}
%%%%%%%%%%%%%%%%%%%%%%%%%%%%%%%%%%%%%%%%%%%%%%%%
%%%%%%%%%%%%%%%%%%%%%%%%%%%%%%%%%%%%%%%%%%%%%%%%

In figure \ref{fig:fig2} (top), we show the onset strain, $\gamma_*$ in the limit cycles for different values of $\lambda$.
We plot both $\gamma_*^+$ in the forward sense (crosses) and $-\gamma_*^-$ in the reverse sense (pluses).
Below $\Gamma_0$ (the vertical lines), the systems are in the elastic regime where the limit cycles are trivial with no plasticity or energy dissipation and $\gamma_*=\Gamma$ in this case.
At $\Gamma_0$, the systems are maximally hardened.
Above $\Gamma_0$, $\gamma_*$ follows an almost perfect line with $\gamma_*=\Gamma_0-(\Gamma-\Gamma_0)$.
It is apparent that for a fixed $\Gamma$, $\gamma_*$ decreases with increasing $\lambda$.
What this means is that at a given $\Gamma$, systems with higher $\lambda$, which are less compressible, exhibit plasticity earlier in the cycle at smaller values of strain.
% We see that systems with higher $\lambda$ yield first with a lower $\Gamma_0$ than systems with a lower $\lambda$ but because the relation $\gamma_*=\Gamma_0-(\Gamma-\Gamma_0)$ has a slope of -1, independent of $\lambda$, systems with higher $\lambda$ also have a smaller $\gamma_*$ for a given $\Gamma$ in the RPR.

In figure \ref{fig:fig2} (middle), we show the stress at the forward plasticity onset (crosses),$\sigma_*^{+}$, and the negative stress at the backward onset (pluses),$-\sigma_*^{-}$, along with their average (dots). 
Similarly, in figure \ref{fig:fig2} (bottom), we plot the plastic strain plateau value at the forward turning point (pluses), $\epsilon_p^{+}$, and the negative plateau value at the reverse turning point (pluses), $-\epsilon_p^{-}$, along with their average (dots).
The asymmetry we see in $\sigma_*$ and $\epsilon_p$ in the middle and bottom rows, respectively, is the result of our choice to begin the simulation by pushing the system in the forward sense but it goes away for larger $\Gamma$.
Similar to $\gamma_*$, in the RPR, $\sigma_*$ decreases with $\Gamma$ with $\frac{d\sigma_*}{d\Gamma}$ approaching a slope of $-1/2$ while $\epsilon_p$ increases from zero with $\frac{d\epsilon_p}{d\Gamma}$ approaching a slope of $1/2$.
We will explain the connection between the decrease in the yield stress and the increase in the hysteresis amplitude below in section~\ref{sec:discussion}, where we introduce a simplified, idealized description of the hardening behavior assuming a constant plastic strain rate and associated perfectly rhombic $\epsilon$ vs. $\gamma$ and $\sigma$ vs. $\gamma$ curves.
%We explain this behavior by taking the derivative of $\sigma=\gamma-\epsilon$ at the onset with respect to $\Gamma$ and requiring that $\frac{d\gamma_*}{d\Gamma}=-1$, from our observation in figure \ref{fig:fig2} (top), then one would get $\frac{d\sigma_*}{d\Gamma}+\frac{d\epsilon_p}{d\Gamma}=-1$. 
%\DV{Message/argument unclear to me}
%Notice that we associate the plastic strain plateau value at the reverse turning point, $\epsilon_p^-$, with the forward $\sigma_*$ and $\gamma_*$ and in that we ignore the creeping regime before the onset and approximate the hysteresis curves in figure \ref{fig:hysteresis} (top) as perfect parallelograms.
%Before $\Gamma_0$, there is no plasticity and $\epsilon_p^{+}=\epsilon_p^{-}>0$.
%This positive residual value is the result of the arbitrary choice to shear the virgin state in the forward direction first.  

%%%%%%%%%%%%%%%%%%%%%%%%%%%%%%%%%%%%%%%%%%%%%%%%
%%%%%%%%%%%%%%%%%%%%%%%%%%%%%%%%%%%%%%%%%%%%%%%%
\begin{figure}[h!]
\includegraphics[width = .7\columnwidth]{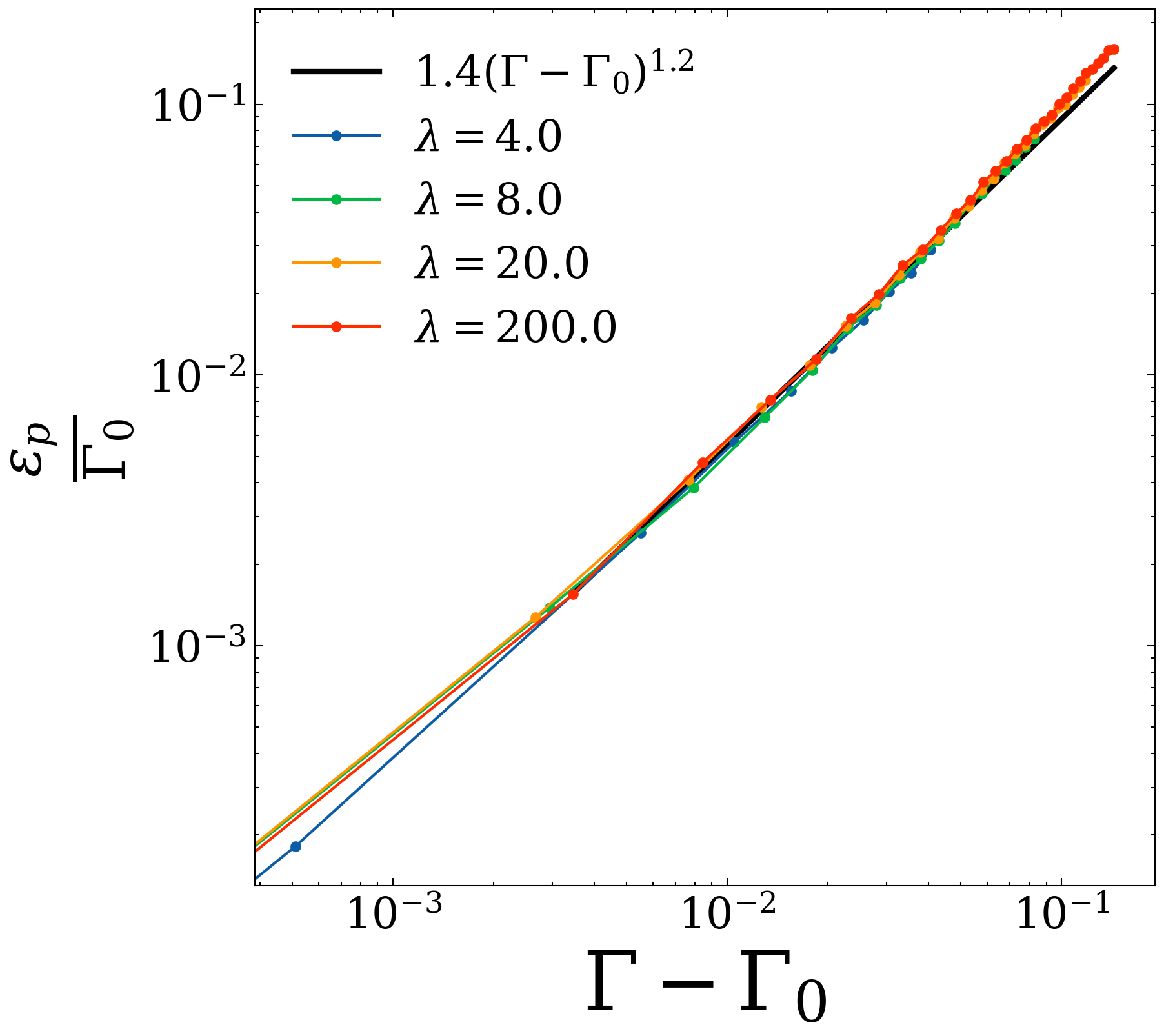}%
\caption{The average of the forward and backward plateau strain $\epsilon_p$ vs. $\Gamma-\Gamma_0$ for different values of $\lambda$.}
\label{fig:eps_p}
\end{figure}
%%%%%%%%%%%%%%%%%%%%%%%%%%%%%%%%%%%%%%%%%%%%%%%%
%%%%%%%%%%%%%%%%%%%%%%%%%%%%%%%%%%%%%%%%%%%%%%%%
In figure \ref{fig:eps_p}, we plot the average of the forward and the backward values of plateau plastic strain, $\epsilon_p$, as in figure \ref{fig:fig2} (bottom), but here we plot $\epsilon_p/\Gamma_0$ vs. $\Gamma-\Gamma_0$.
The data collapse almost perfectly onto a single master curve.
The master curve is reasonably well represented by a power law with an exponent of $1.2$ and a prefactor of $1.4$ which is consistent with our result for the incompressible case \cite{elgailani2022anomalous}.
It is not immediately obvious why one would need to scale $\epsilon_p$ by $\Gamma_0$, but, empirically, it is necessary to obtain a good collapse. 

\subsection{Stress distribution}
%%%%%%%%%%%%%%%%%%%%%%%%%%%%%%%%%%%%%%%%%%%%%%%%
%%%%%%%%%%%%%%%%%%%%%%%%%%%%%%%%%%%%%%%%%%%%%%%%
\begin{figure*}
\includegraphics[width = 2.\columnwidth]{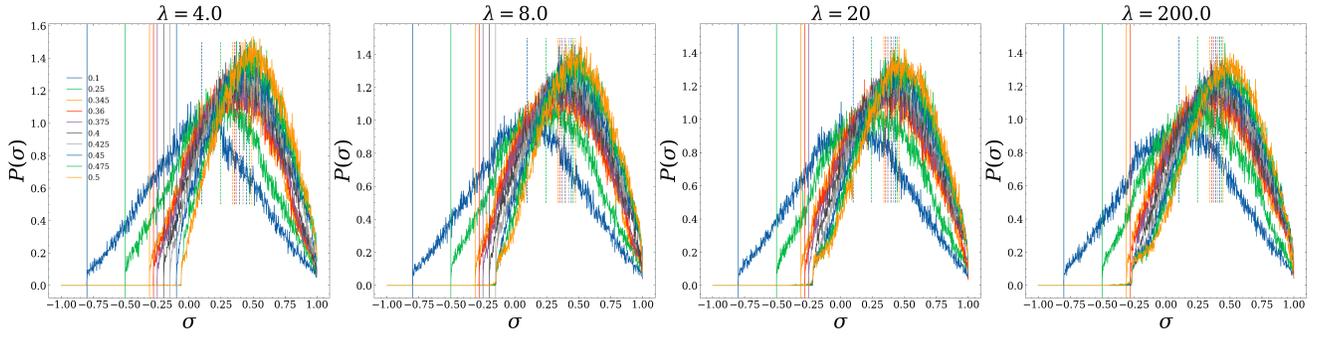}
\caption{The distribution of the stress $\sigma$ at the forward turning point ($\gamma=\Gamma$) in the terminal limit cycles for different cycling amplitudes, $\Gamma$.
%Data are averaged over all $T$ configurations in a period $T$ limit cycle.
The vertical dashed lines are the averages.
The vertical solid lines are at the cutoff values of $\sigma=2\Gamma-1$, below which there can be no weight for terminally elastic systems in the ER.
%\DV{Test here a table $2\times 2$ instead of $4\times 1$.}
}
\label{fig:fig4}
\end{figure*}
%%%%%%%%%%%%%%%%%%%%%%%%%%%%%%%%%%%%%%%%%%%%%%%%
%%%%%%%%%%%%%%%%%%%%%%%%%%%%%%%%%%%%%%%%%%%%%%%%%%%%%%%%%%%%%%%%%
 In figure \ref{fig:fig4} we plot the distribution of the local stress, $P(\sigma)$, at the forward turning point for several different values of $\Gamma$ for the same set of $\lambda$ as above.
 The vertical dashed lines are at the average.
 The vertical solid lines are located at $-1+2\Gamma$ and shown for only the elastic systems.
 
 For systems in the ER, the distribution is symmetric about the average (dashed lines) which sits at a positive value because we show the forward turning point where the stress is positive.
 Because there is no plasticity for the systems in the ER as we strain them backward from $+\Gamma$ to $-\Gamma$, the distributions will simply shift horizontally to the left by $2\Gamma$.
 This means that at the forward turning points, $P(\sigma)$ must be zero for $\sigma<-1+2\Gamma$ as indicated by the vertical solid lines.
 Otherwise, if $P(\sigma)$ was non-zero for some $\sigma<-1+2\Gamma$ then the system would exhibit some plasticity during reverse shear and therefore could not be elastic.

For all systems, both the elastic systems below $\Gamma_0$ and the reversible plastic systems above $\Gamma_0$, a feature appears at $\sigma_0$.
%For the elastic systems, the distribution has an excess above the background at both $\sigma_0$ and $2\Gamma-\sigma_0$.
For the elastic systems, the distribution has an excess above the background around $\sigma_0$.
The excess sites with $\sigma\approx\sigma_0$ presumably result from the last few forward strain sweeps during the transient before plasticity was extinguished.
During these sweeps, the sites which yielded just before the reversal point would have ended up with a final strain value of $\sigma_0$.
%The excess sites at $\sigma=2\Gamma-\sigma_0$ were similarly produced during the final few backward strain sweeps. 

The distributions for systems in the RPR are asymmetric.
%Where the elastic systems had a bump, the plastic systems develop a shoulder at $\sigma_0$.
Precisely at the transition point, $\Gamma=\Gamma_0$, $P(\sigma)$ remains roughly symmetric about the average and the peak remains roughly at the average. 
However, as $\Gamma$ increases, the stress at the turning point -- i.e., the average $\sigma$ -- increases, but it does so more slowly with $\Gamma$ than the location of the peak, and the distributions become increasingly asymmetric further above the transition.  
%As $\Gamma$ increases above $\Gamma_0$, a shoulder develops at $\sigma_0$.
Precisely at the transition, $P(\sigma)$ is still zero to the left of $\sigma_0$ as in the elastic case, but as $\Gamma$ increases, the region below $\sigma_0$ becomes increasingly populated while a shoulder develops immediately to the right of $\sigma_0$.
% The maximum of $P(\sigma)$ is significantly to the right of the average.
The height of the jump at the shoulder is $\approx0.1$ regardless of $\Gamma$ or $\lambda$.
For systems just above the transition with $\Gamma$ just greater than $\Gamma_0$, there is hardly any mass of $P(\sigma)$ at $\sigma<\sigma_0$ and the jump is relatively sharp.
As $\Gamma$ increases beyond $\Gamma_0$ the mass of $P(\sigma)$ at $\sigma<\sigma_0$ slightly increases, smoothing the jump.
%{\bf XXX-AE: repeated point?} The explanation for the emergence of the shoulder is that in our model, when a site hits threshold at $\sigma=1$, its stress value transforms from $1$ to $\sigma_0$.

% The $P(\sigma)$ distributions explain the dependence of $\gamma_*$ on $\Gamma$ and the emergence of $\Gamma_0$.
The $P(\sigma)$ distributions explain the origin of $\gamma_*$ and $\Gamma_0$.
For the moment, if we ignore the very small mass of $P(\sigma)$ in the creeping regime at $\sigma<\sigma_0$, then, as we load the systems in the reverse direction, $P(\sigma)$ will simply shift to the left until the shoulder reaches the reverse yielding point at $\sigma=-1$.
This will occur after a reverse straining by a magnitude of $(1+\sigma_0)$.
Since we start the reverse strain from the turning point at $\gamma=\Gamma$, the value of the strain at the onset of yielding in the reverse direction will occur at $\gamma=\Gamma-(1+\sigma_0)$.
Since this is the onset of plasticity in the reverse direction, we must have that $\gamma_*=(1+\sigma_0)-\Gamma$ for systems in the RPR.
At the critical point, $\Gamma=\Gamma_0$, the onset of plasticity occurs just at the backward turning point, so we must have: 
\begin{equation}
 \Gamma_0=\frac{1+\sigma_0}{2}=1-\alpha A_0-\frac{1}{L^2}.
 \label{eq:Gamma0}
\end{equation}
Therefore 
\begin{equation}
\gamma_*=\Gamma_0-(\Gamma-\Gamma_0).
\label{eq:gamma*}
\end{equation}

%%%%%%%%%%%%%%%%%%%%%%%%%%%%%%%%%%%%%%%%%%%%%%%%
%%%%%%%%%%%%%%%%%%%%%%%%%%%%%%%%%%%%%%%%%%%%%%%%
\begin{figure}[h!]
\includegraphics[width = 0.7\columnwidth]{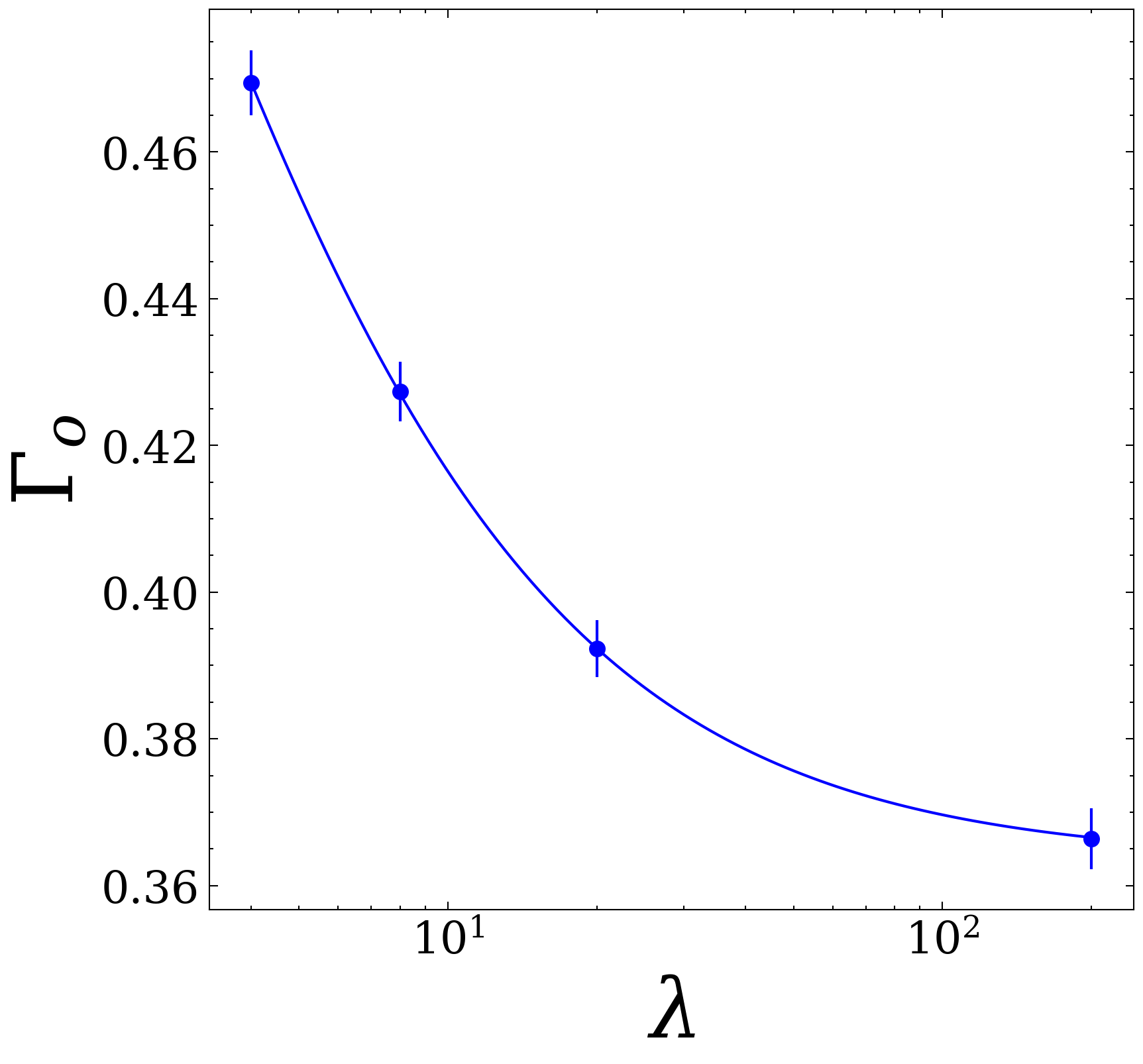}%
\caption{The transition amplitude, $\Gamma_0$, from the elastic regime (ER) to the reversible plastic regime (RPR) vs. Lame parameter $\lambda$.
% $\sigma_y$ is obtained by averaging the stress values (sampled at equal intervals of strain $\Delta\gamma=5E-3$) after a sufficiently large strain at $\gamma = 4.0$.
The solid line is the prediction from the Eshelby kernel using the relation $\Gamma_0=(1+\sigma_0)/2$ (equation~\ref{eq:Gamma0})
The dots are measured by averaging $(\Gamma+\gamma_*)/2$ in figure \ref{fig:fig2} (top) for $\Gamma>\Gamma_0$.
The error bars are the standard error of the mean. The dashed line is at the incompressible limit of $\Gamma_0$.}
\label{fig:Gamma0}
\end{figure}
%%%%%%%%%%%%%%%%%%%%%%%%%%%%%%%%%%%%%%%%%%%%%%%%
%%%%%%%%%%%%%%%%%%%%%%%%%%%%%%%%%%%%%%%%%%%%%%%%

In figure \ref{fig:Gamma0}, we show the ER to RPR transition amplitude, $\Gamma_0$, vs. $\lambda$.
The dots are the values measured from $\gamma_*$ vs. $\Gamma$ curves in the RPR in figure \ref{fig:fig2} (top) by taking the average of $(\gamma_*+\Gamma)/2$.
The solid line is our prediction from the Eshelby kernel using the relation $\Gamma_0=(1+\sigma_0)/2$.
Independent of $\lambda$, the two values match to within symbol size error. 
As we increase $\lambda$, $\Gamma_0$  approaches the incompressible limit at $0.36320$.

%%%%%%%%%%%%%%%%%%%%%%%%%%%%%%%%%%%%%%%%%%%%%%%%
%%%%%%%%%%%%%%%%%%%%%%%%%%%%%%%%%%%%%%%%%%%%%%%%
\begin{figure}
\includegraphics[width = .466\columnwidth]{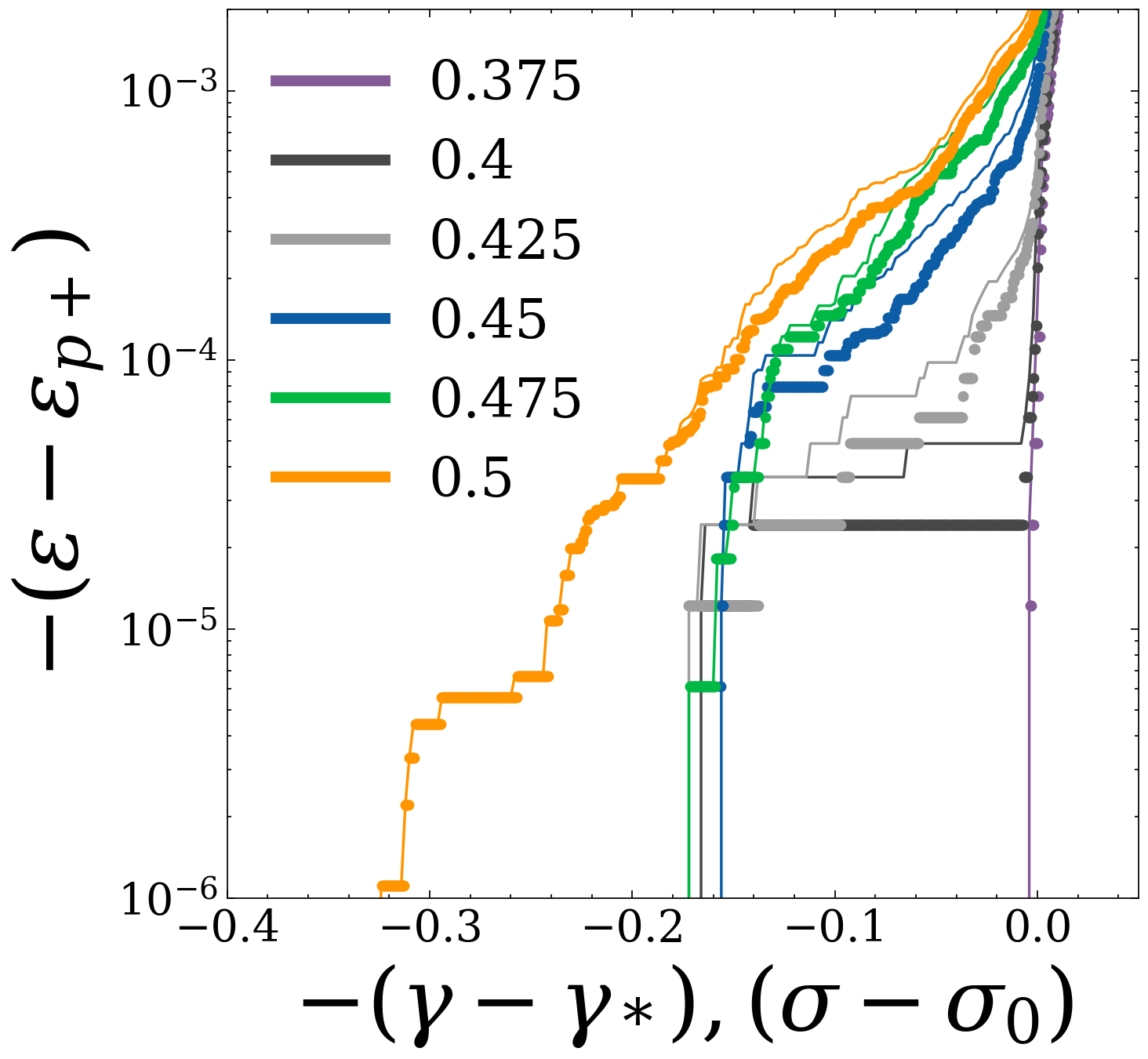}
\includegraphics[width = .5\columnwidth]{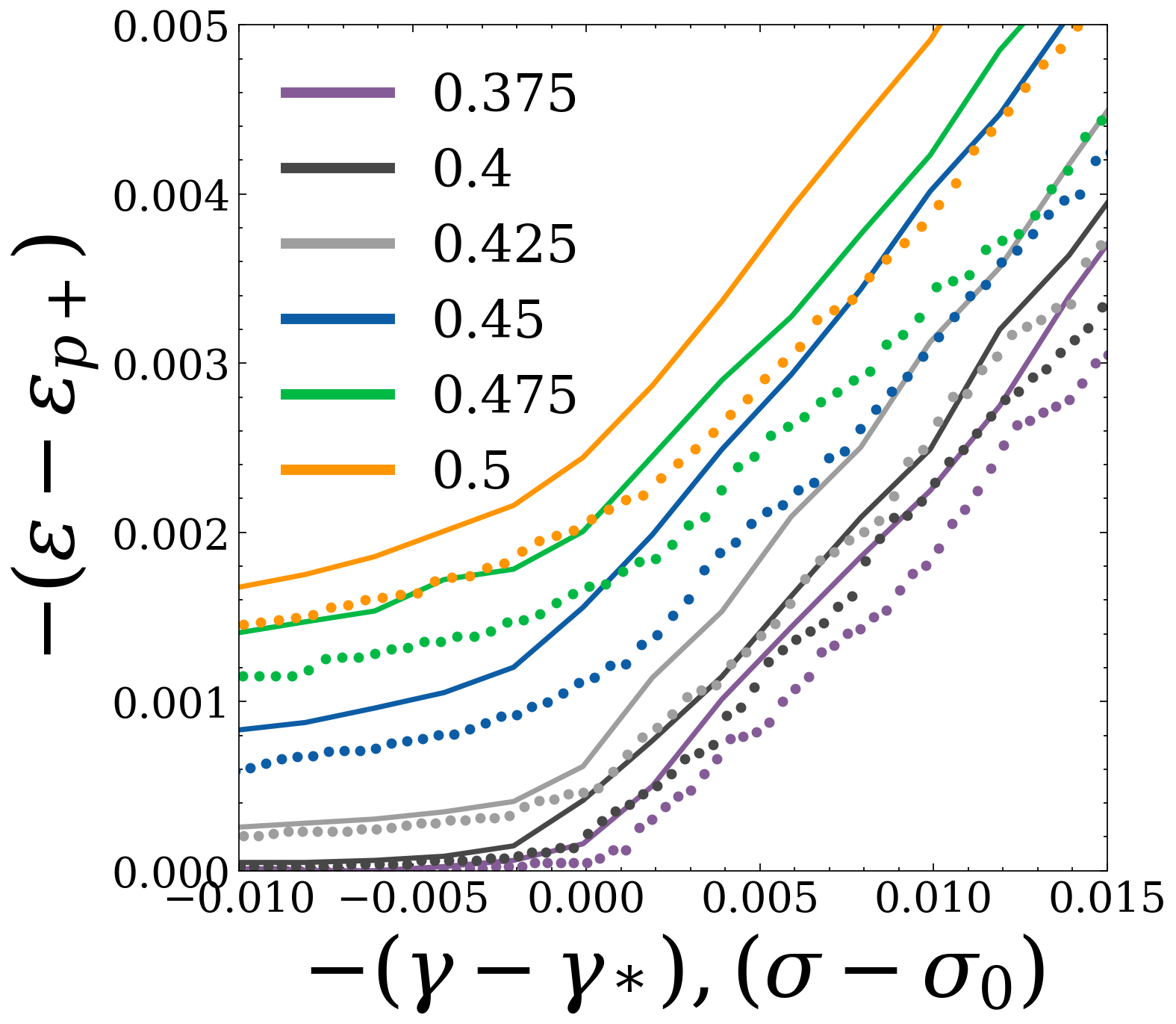}
%\begin{center}
%\includegraphics[width = .38\textwidth]{plastic-creep-200.0.png}
%\hspace{1cm}
%\includegraphics[width = .405\textwidth]{plastic-creep-200.0-2.png}
%\end{center}
\caption{The plastic strain, $\epsilon$, vs. $\gamma-\gamma_*$ and $\sigma-\sigma_0$ for $\lambda=200$.
The dots are the measured values of $\epsilon$ as in figure \ref{fig:hysteresis} but with the forward plateau value $\epsilon^+$ subtracted off.
The solid lines are estimates of $\epsilon$ obtained from $P(\sigma)$ at the forward turning points as $2/L^2\int_{-\infty}^{\sigma} P(\sigma^\prime)d\sigma^\prime$.
The two plots show the same data with different scales and ranges of the axes.
%\DV{test with increased size of graphs }
}
\label{fig:creep}
\end{figure}
%%%%%%%%%%%%%%%%%%%%%%%%%%%%%%%%%%%%%%%%%%%%%%%%
%%%%%%%%%%%%%%%%%%%%%%%%%%%%%%%%%%%%%%%%%%%%%%%%%%%%%%%%%%%%%%%%%
% In figure \ref{fig:creep}, we plot the plastic strain, $\epsilon$, shifted by $-\epsion_{p+}$, vs. $\gamma-\gamma_*$ around the backward turning point for $\lambda=200$.
In figure \ref{fig:creep}, for the $\lambda=200$ ensemble, we plot the plastic strain accumulated during a backward strain sweep from $\gamma=+\Gamma$ to $\gamma=-\Gamma$.
The horizontal axis, $\gamma-\gamma_*$, stands for the magnitude of strain, $\delta\gamma$
% So the y-axis, $\epsilon-\epsion_{p+}$, is the accumulated plasticity as we shear the systems from $\gamma=+\Gamma$ to $\gamma=-\Gamma$.
The dotted lines are the measured $\epsilon$ as in figure \ref{fig:hysteresis} (top).
The solid lines are estimates of $\epsilon-\epsilon_{p+}$ obtained from $P(\sigma)$ by $(2/L^2)C(\sigma-\sigma_0)$ where $C$ is the cumulative distribution function $C(\sigma)=\int_{-\infty}^{\sigma} P(\sigma^\prime)d\sigma^\prime$.
The prefactor of $2/L^2$ is the plastic increment resulting from a single flip.
Here we neglect the stress redistribution after plastic events, nevertheless, we show that this approximation qualitatively captures the main features of the loading curves. 
A similar approach was taken in reference~\cite{PhysRevLett.124.205503} to synthetically generate forward and backward loading curves to study mechanical polarization in Lennard-Jones glasses.  
%\DV{check mean field like argument in the Bauschinger effect paper - Patinet et al PRL 124, 205503 (2020) around eq. (2)}
% In figure \ref{fig:creep} (left), we plot the data on a semi-log scale in the creeping regime for $\gamma<\gamma_*$.
% As discussed above, 

Both the directly measured $\epsilon$ curves and the ones inferred from $P(\sigma)$ show a sharp transition at $\gamma_*$ that becomes more and more rounded as $\Gamma$ gets further away from $\Gamma_0$.
For instance, the system at $\Gamma=0.375$ (purple) has essentially no creep and the slope, $\frac{d\epsilon}{d\gamma}$, jumps from zero to a finite value at $\gamma_*$, while for the system at $\Gamma=0.5$ the transition is smoother.
Nevertheless, the slopes above $\gamma_*$ are relatively insensitive to $\Gamma$.
Furthermore, for any $\gamma-\gamma_*$, there is more creep the larger $\Gamma-\Gamma_0$.
% The information from $C(\sigma)$ qualitatively captures the main features of the loading curves.
The slopes, $\frac{d\epsilon}{d\gamma}$, of the actual data and the ones inferred from $P(\sigma)$ are roughly equal. 
They all start from essentially zero to the left of $\gamma_*$ and jump to a finite value of $0.2$ to the right of $\gamma_*$.

% The excess of $P(\sigma)$ at $\sigma<\sigma_0$ corresponds to the creeping regime in $\epsilon$ for $\gamma<\gamma_*$  we see in figure \ref{fig:hysteresis}(middle).
% When a site transforms it increments $\epsilon$ by $2/L^2$ if we ignore the stress redistribution associated with the transformation which can possibly trigger other sites near threshold.

% While, theoretically, $\lambda$ can be any real number we focus on $\lambda>0$ in this study.

\section{Forward shearing}
\label{sec:forward-shear}

 %%%%%%%%%%%%%%%%%%%%%%%%%%%%%%%%%%%%%%%%%%%%%%%%
%%%%%%%%%%%%%%%%%%%%%%%%%%%%%%%%%%%%%%%%%%%%%%%%
\begin{figure}[h!]
\includegraphics[width = .475\columnwidth]{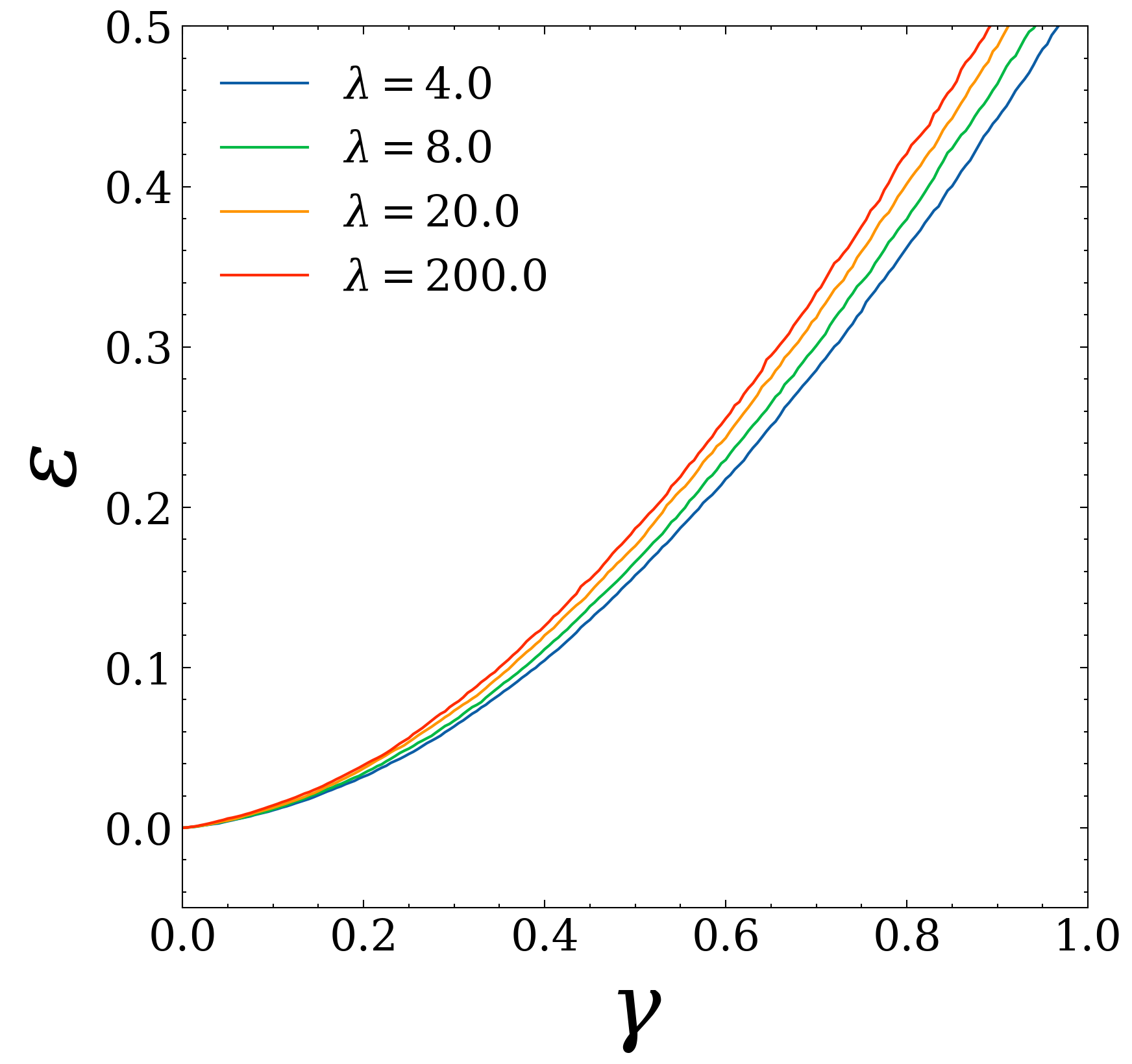}
\includegraphics[width = .495\columnwidth]{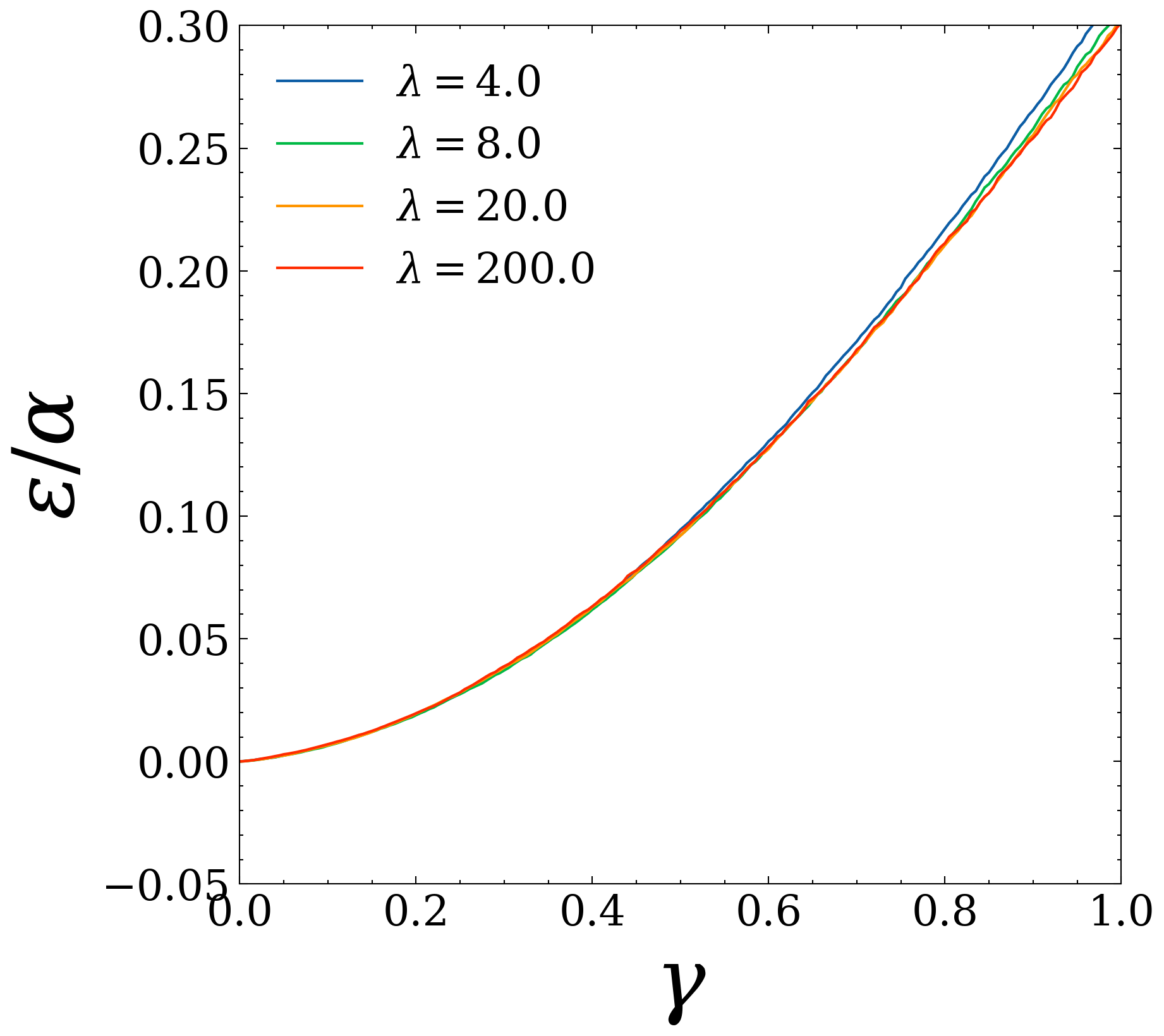}
\caption{The plastic strain (ensemble average), $\epsilon$, vs strain, $\gamma$, in forward shear for different values of $\lambda$ . We subtract off the initial value of $\epsilon$ which is a residual effect of the random quench and scales like $1/L^2$ divided by the number of systems in the ensemble,
$s$ all curves begin precisely at zero. The values are sampled at equal
increments of strain $\Delta \gamma = 5 \times 10^{-3}$. Right: same curves as on the left
after we scale   $\epsilon$ by $\alpha$.
}
\label{fig:fwd_shear_epsilon}
\end{figure}
In figure~\ref{fig:fwd_shear_epsilon}, we plot the plastic strain vs. applied strain in forward steady shearing.
On the left, we plot the unscaled data.
We see that the systems with larger $\lambda$ have higher plastic strain at a given applied total strain.
On the right, we scale the plastic strain by the $\alpha$ parameter which controls the amplitude of the stress \emph{redistribution} in the model.
The collapse is quite good in the early stages of shearing which shows that the number of shear transformations in a given amount of applied strain is proportional to $\alpha$.
However, the collapse becomes less accurate at larger strains as the systems begin to approach steady-state shear.

%%%%%%%%%%%%%%%%%%%%%%%%%%%%%%%%%%%%%%%%%%%%%%%%
%%%%%%%%%%%%%%%%%%%%%%%%%%%%%%%%%%%%%%%%%%%%%%%%
\begin{figure}[h!]
\includegraphics[width = .49\columnwidth]{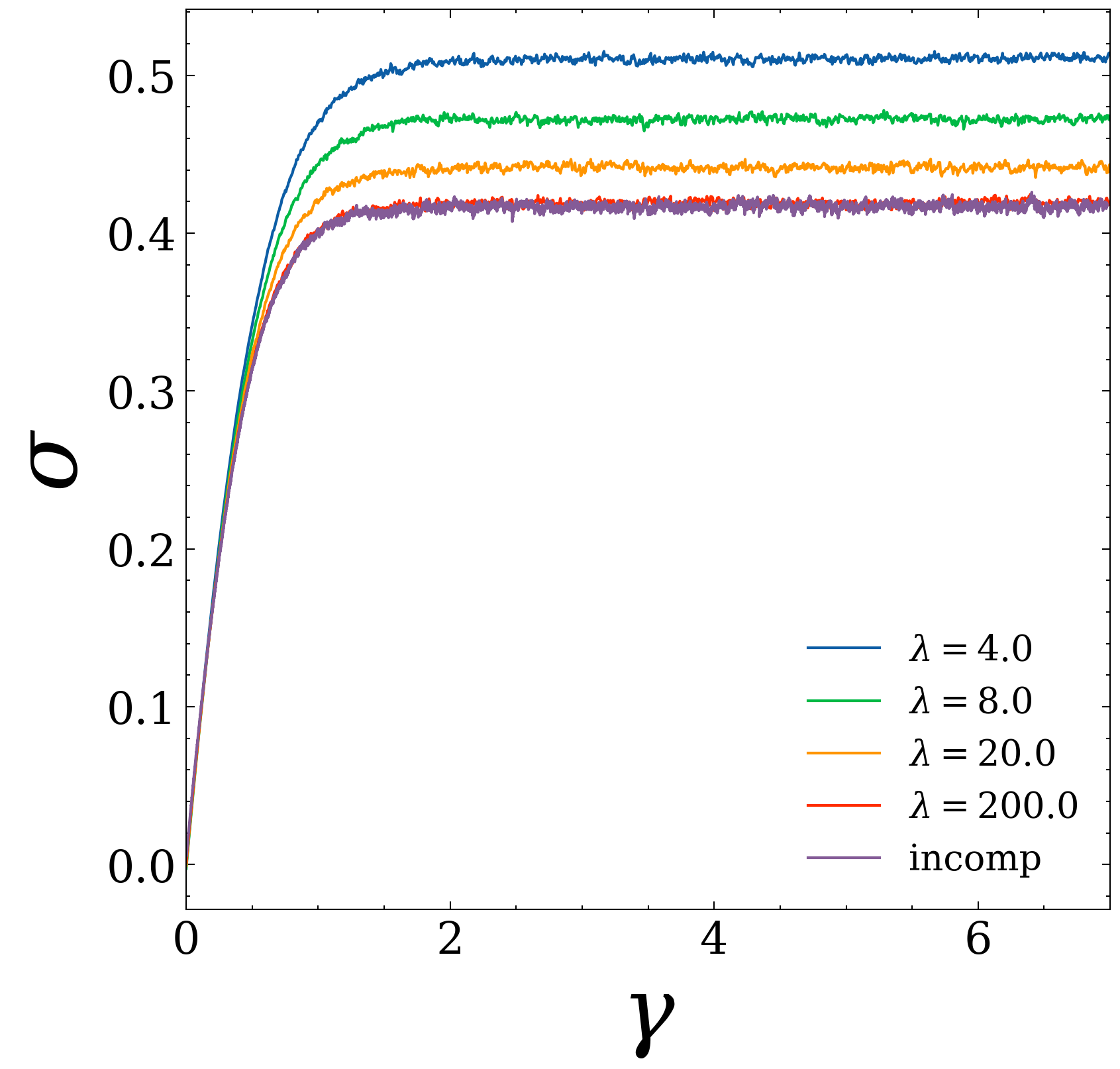}
\includegraphics[width = .49\columnwidth]{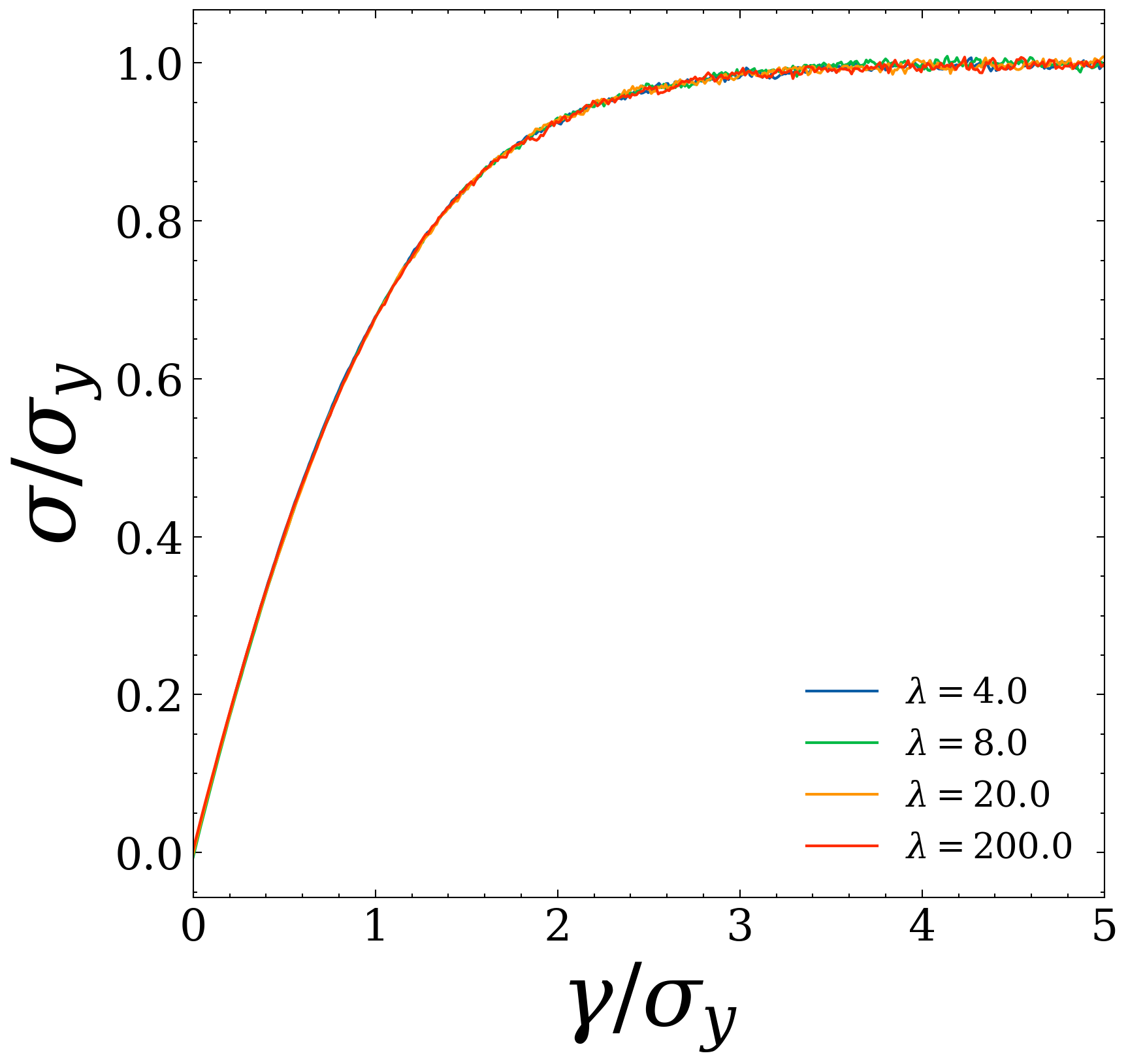}

\caption{Left: The shear stress (ensemble average) $\sigma$ vs strain $\gamma$ in forward shear for different values of $\lambda$ along with the incompressible limit (purple).
The stress values are sampled at equal increments of strain $\Delta\gamma=5\times 10^{-3}$.
% The inset shows the stress rate $\frac{d\sigma}{d\gamma}$ vs. $\gamma$.
% The vertical and horizontal lines are at the corresponding transition amplitudes in cyclic shearing $\Gamma_0$.
Right:  same curves as on the left after we normalize both $\sigma$ and $\gamma$ by the flow stress, $\sigma_y$ in forward shear.
% Right:  The shear stress, $\sigma$, normalized by the flow stress, $\sigma_y$, vs. strain, $\gamma$, normalized by the flow stress, $\sigma_y$ in forward shear
}
\label{fig:fwd_shear}
\end{figure}
%%%%%%%%%%%%%%%%%%%%%%%%%%%%%%%%%%%%%%%%%%%%%%%%
%%%%%%%%%%%%%%%%%%%%%%%%%%%%%%%%%%%%%%%%%%%%%%%%
In figure \ref{fig:fwd_shear} (left), we plot the shear stress (ensemble average)\footnote{
%Note that avalanches are present and they occur at fixed values of strain. The stress drops during avalanches not shown in the curves since avalanches are not the focus of this study and we are basically sampling the stress between the avalanches.
Note that large stress drops are present in the automaton dynamics at particular strain values, but we sample the stress and plastic strain at regular strain intervals, so the avalanches are not visible in the stress vs. strain curves.
}, $\sigma$,  vs. the total strain $\gamma$ in forward shear.
%We show the curves for different values of $\lambda$ and we also show the incompressible case. 
All curves eventually approach a steady state flow stress, $\sigma_y$, with a slope of zero.
$\sigma_y$ decreases monotonically with increasing $\lambda$.
In figure \ref{fig:fwd_shear} (right), we plot the same curves as in figure \ref{fig:fwd_shear} (right) but here we normalize both $\sigma$ and $\gamma$ by $\sigma_y$.
The rationale for scaling $\sigma$ by $\sigma_y$ is obvious; the rationale for scaling $\gamma$ by $\sigma_y$, is that we know the curves all start with a slope near unity as the plastic strain rate is small initially, and in order to maintain the same slope of unity in all curves, $\gamma$ must be rescaled by the same amount as $\sigma$ for any hope of a collapse.
Indeed, the curves do collapse remarkably onto a single master curve after the re-scaling.
We note that the good collapse of the plastic strain in figure~\ref{fig:fwd_shear_epsilon} after scaling $\epsilon$ by $\alpha$ but leaving $\gamma$ unscaled implies that if we were to scrutinize the collapse of the stress in figure~\ref{fig:fwd_shear}, we would see a better collapse using $\alpha$ in the early stages rather than $\sigma_y$.

%%%%%%%%%%%%%%%%%%%%%%%%%%%%%%%%%%%%%%%%%%%%%%%%
%%%%%%%%%%%%%%%%%%%%%%%%%%%%%%%%%%%%%%%%%%%%%%%%
\begin{figure}[h!]
\includegraphics[width = .49\columnwidth]{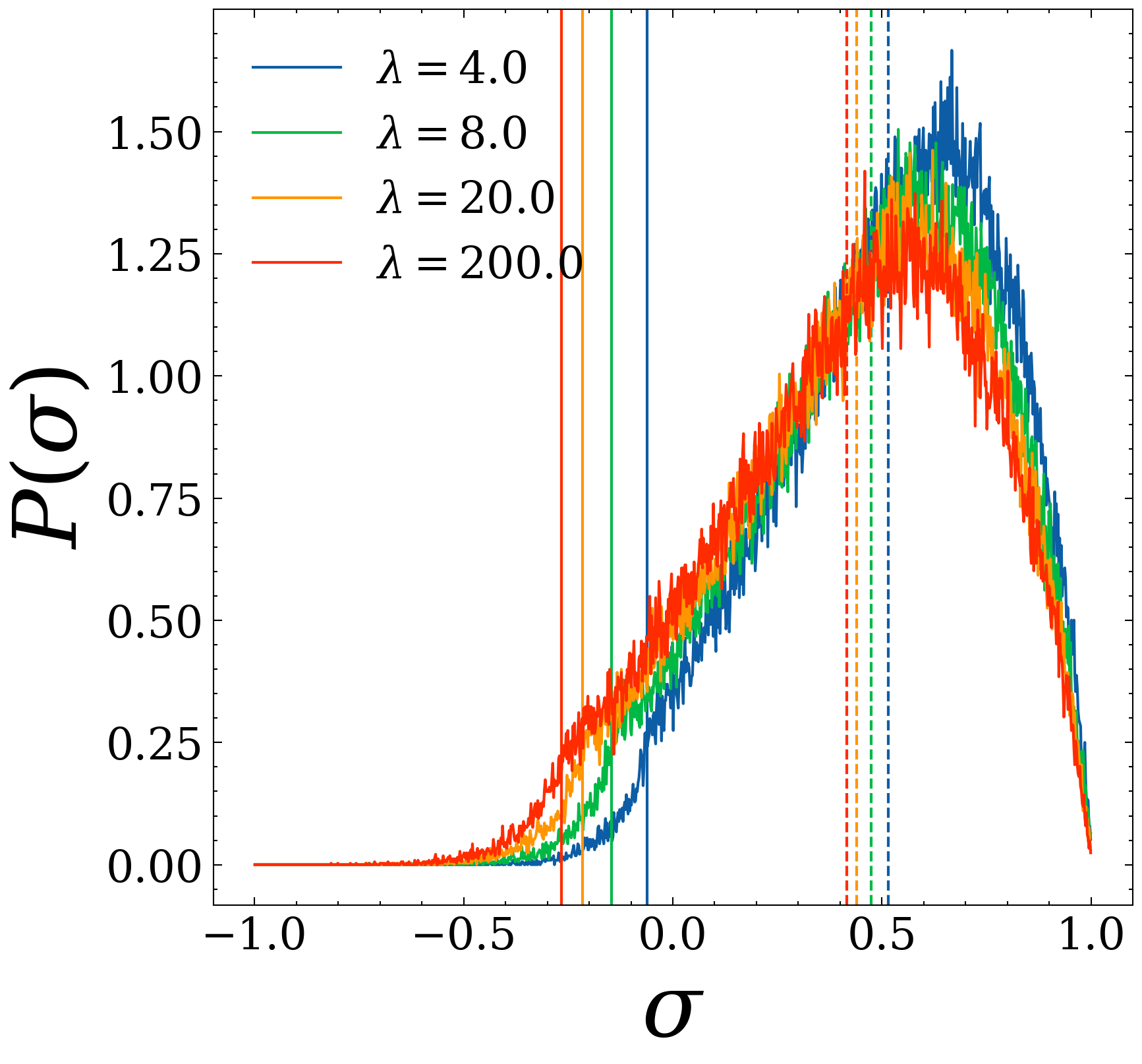}
\includegraphics[width = .49\columnwidth]{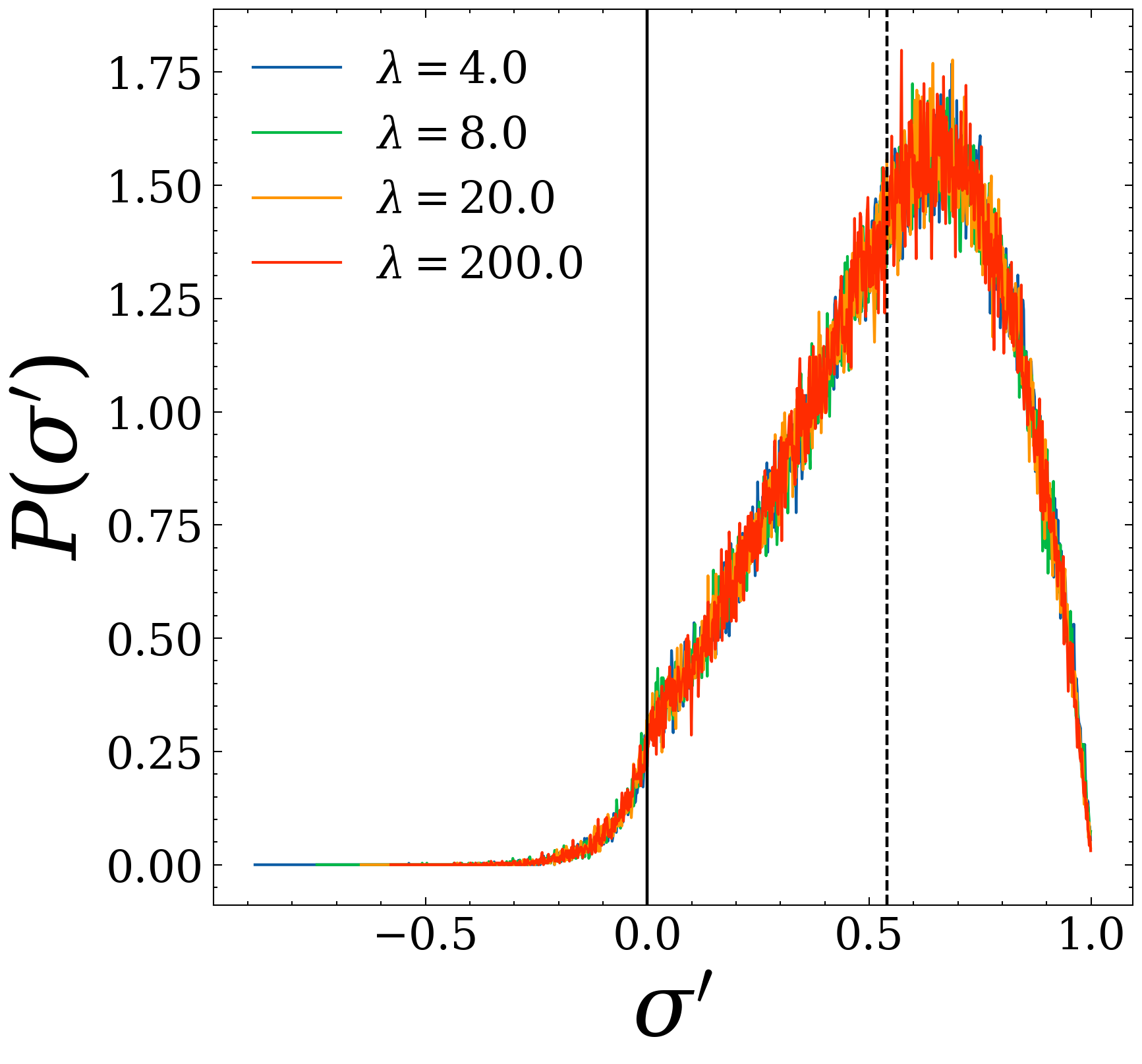}
\caption{Left: the distribution of the shear stress $P(\sigma)$ in the steady state of forward shearing for different values of $\lambda$.
Right: same data on the left but here we perform a coordinate transformation on the stress: $\sigma\to\sigma^{\prime}=\frac{\sigma-\sigma_0}{1-\sigma_0}$.
The vertical solid line is at $\sigma^\prime_0=0$.
The vertical dashed line is the average of $\sigma^\prime$ over the four distributions: $\sigma^\prime_y=0.54014$.}
% Right: the distribution of the shear stress $P(\sigma^{\prime})$ in the stead state where $\sigma^{\prime}=\frac{\sigma-\sigma}{1-\sigma_0}$.}
\label{fig:fwd_shear_2}
\end{figure}
%%%%%%%%%%%%%%%%%%%%%%%%%%%%%%%%%%%%%%%%%%%%%%%%
%%%%%%%%%%%%%%%%%%%%%%%%%%%%%%%%%%%%%%%%%%%%%%%%
In figure \ref{fig:fwd_shear_2} (left), we show the shear stress distribution $P(\sigma)$ in the steady state of forward shearing.
The solid lines are at $\sigma_0$.
The dashed lines are at the averages.
The peaks are located at significantly higher sigma than the averages.
$P(\sigma)$ gets increasingly broad with increasing $\lambda$ pushing both the peak and the average $\sigma$ to lower values as $\sigma_0$ becomes increasingly negative.
Reminiscent of the case of cyclic shearing, $P(\sigma)$ has a shoulder around $\sigma_0$ resulting from the fact that transforming sites re-enter the distribution precisely at $\sigma_0$. 

Motivated by the similar form of the distribution at the shoulder, we make a transformation to a new variable, $\sigma^{\prime}$, where
\begin{equation}
 \sigma^{\prime}=\frac{\sigma-\sigma_0}{1-\sigma_0}
 \label{eq:sigmaPrime}
\end{equation}
in order to fix the location of the shoulder to $\sigma^{\prime}=0$ while leaving the elementary yielding threshold at $1$.
In figure \ref{fig:fwd_shear_2} (right), we plot $P(\sigma^{\prime})$.
After this coordinate transformation, the different distributions in figure \ref{fig:fwd_shear_2} (right) are virtually indistinguishable.

In particular, this rescaling tells us that the emergent macroscopic yield stress, $\sigma_y$, has a simple dependence on $\lambda$.
We find that the average $\sigma^\prime$ value, $\sigma_y^\prime$,  is essentially independent of $\lambda$ and equal to $0.54 \pm 0.002$ (when averaged over our $4$ different $\lambda$ values).
This tells us that the yield stress must have the form:
\begin{eqnarray}
\sigma_y&=&\sigma_y^\prime (1-\sigma_0)+\sigma_0 \\
&=& \sigma_y^\prime + \sigma_0 (1-\sigma_y^\prime)\\
&=& \sigma_y^\prime + [1-2\alpha A_0 - \frac{2}{L^2}][1-\sigma_y^\prime]\label{eq:sigmay}
\end{eqnarray}

%%%%%%%%%%%%%%%%%%%%%%%%%%%%%%%%%%%%%%%%%%%%%%%%
%%%%%%%%%%%%%%%%%%%%%%%%%%%%%%%%%%%%%%%%%%%%%%%%
\begin{figure}[h!]
\includegraphics[width = .7\columnwidth]{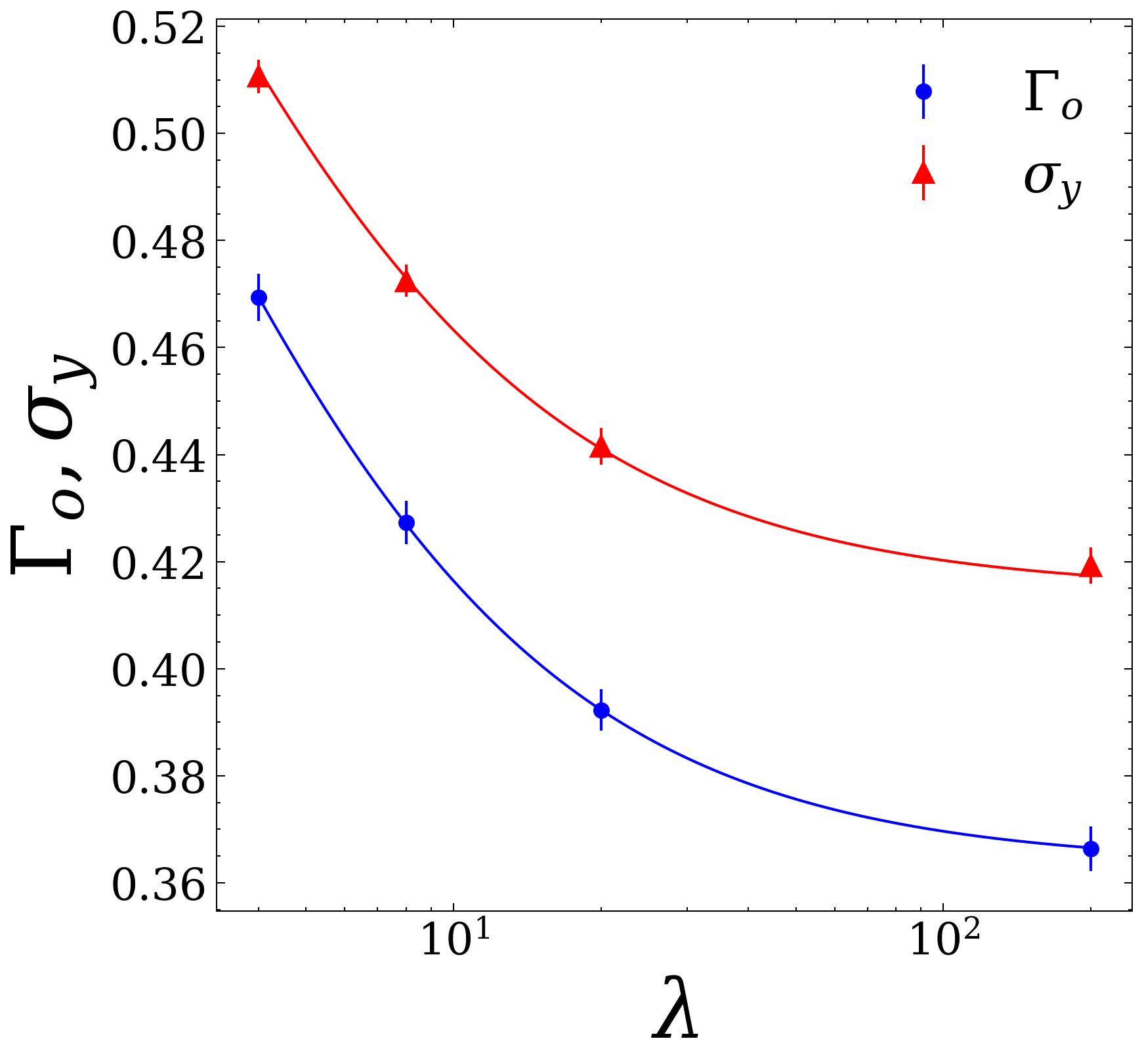}
\caption{
The flow stress, $\sigma_y$, in forward shearing vs. Lame first parameter $\lambda$ (triangles).
We also include $\Gamma_0$ reproduced from figure \ref{fig:Gamma0} for comparison (dots).
The red solid line is $\sigma_y$ calculated from equation~\ref{eq:sigmay}
% The red triangles are measurements of $\sigma_y$ (ensemble average) obtained by averaging the stress values (sampled at equal intervals of strain $\Delta\gamma=5E-3$) after a sufficiently large strain  $\gamma = 2.0$ in forward shearing.
}
\label{fig:sigmay}
\end{figure}
%%%%%%%%%%%%%%%%%%%%%%%%%%%%%%%%%%%%%%%%%%%%%%%%
%%%%%%%%%%%%%%%%%%%%%%%%%%%%%%%%%%%%%%%%%%%%%%%%

In figure \ref{fig:sigmay}, we plot $\sigma_y$ vs. $\lambda$; and we include $\Gamma_0$ from the previous figure for comparison.
The triangles are the averages of $\sigma_y$
measured from the curves in figure \ref{fig:fwd_shear} in the steady state; where we conventionally define the steady state as $\gamma \ge 2$.
The red solid line is $\sigma_y$ calculated by equation~\ref{eq:sigmay} with $\sigma_y^\prime=0.54$.
The two curves have the same trend with $\lambda$; monotonically decreasing with increasing $\lambda$.
$\Gamma_0$ sits below $\sigma_y$ at all $\lambda$ values, consistent with our scaling arguments above.

In table \ref{table1} we list the expressions for $\sigma_c$, $\sigma_0$, $\Gamma_0$ and $\sigma_y$.
Here we ignore terms that are proportional to $\frac{1}{L^2}$ and we set $\mu=1$.
Again, $\sigma_c$ is the constrained stress at the origin of solution to the discretized Eshelby problem; where we impose $\epsilon=1$ at the origin in an initially stress-free material and solve for the displacement field.
$\sigma_0$ is the backstress in our model, which is the shear stress at a tile after the tile reach the stress threshold and transforms.
$\Gamma_0$ is the ER to RPR transition amplitude in cyclic shear.
$\sigma_y$ is the steady state flow shear stress in forward shear.
$A_0=A[0,0]=0.317674$.
Where, again, $\alpha=2(\lambda+\mu)/(\lambda+2\mu)=2K/(K+\mu)$ as defined in equation~\ref{eq:alpha}.

\begingroup
\setlength{\tabcolsep}{10pt} % Default value: 6pt
\renewcommand{\arraystretch}{1.8} % Default value: 1
\begin{table}
\label{table1}
\caption{Summary of $\sigma_c$, $\sigma_0$, $\Gamma_0$ and $\sigma_y$ expressions provided in the text.
Here we ignore terms that are proportional to $\frac{1}{L^2}$.}
\begin{center}
\begin{tabular}{ |c|c| } 
 \hline
 $\sigma_c$ & $\approx -\alpha A_0$ \\ 
  \hline
 $\sigma_0$ & $\approx 1-2\alpha A_0$  \\ 
  \hline
 $\Gamma_0$ & $\approx 1-\alpha A_0$ \\
  \hline
 $\sigma_y$ & $\approx \sigma_y^{\prime}+\left[1-2\alpha A_0\right](1-\sigma_y^{\prime})$ \\
 \hline
\end{tabular}
\end{center}
\end{table}
\endgroup

% \begingroup
% \setlength{\tabcolsep}{10pt} % Default value: 6pt
% \renewcommand{\arraystretch}{1.8} % Default value: 1
% \begin{center}
% \begin{tabular}{ |c|c| } 
%  \hline
%  $\sigma_c$ & $-\frac{0.63666+0.63666\lambda}{\lambda+2}$ \\ 
%  \hline
%  $\sigma_0$ & $\frac{0.726674-0.27320\lambda}{\lambda+2}$  \\
%  \hline
% $\Gamma_0$ & $\frac{1.36334+0.36340\lambda}{\lambda+2}$ \\
% \hline
%  $\sigma_y$ & $\frac{1.41427+0.41432\lambda}{\lambda+2}$ \\
%  \hline
% \end{tabular}
% \end{center}
% \endgroup

%\section{Discussion}
\section{Explanation of emergent hardening behavior}
\label{sec:discussion}

For most materials, plastic behavior depends on the history of past deformations. 
These can tend to harden, soften and polarize the initial material. 
Whether metal, glass or polymer, strain-hardening is intimately linked to the evolution of the microstructure under deformation: dislocation entanglement, local rearrangements of an amorphous structure, polymer chain alignments, and so on.
At the macroscopic level, continuum plasticity models assume ad-hoc hardening laws that prescribe the evolution of the yielding condition -- prescribed by the so-called yield surface -- during deformation in order to accurately account for experimental results~\cite{KahnHuangBook}.

Interestingly, at the mesoscopic scale, hardening appears to be an emerging property of elasto-plastic models. 
In Ref.~\cite{BVR-PRL02,Talamali:2012aa}, the local plastic thresholds that limit the extent of elastic behavior are drawn from a random distribution.
During deformation, it was shown that the gradual exhaustion of weak local plastic thresholds and their replacement by normal thresholds (associated with a new local amorphous structure) induces systematic hardening behavior: the greater the plastic deformation, the higher the macroscopic yield strength. 
A similar mechanism has recently been invoked to explain the development of a Bauschinger effect in model glasses~\cite{PhysRevLett.124.205503}.

In the present model, we identify two distinct hardening regimes. 
As detailed in previous sections and in Ref.~\cite{Khirallah-PRL2021}, for cycling amplitudes $\Gamma < \Gamma_0$, after a transient, the system locks in onto a perfectly elastic limit cycle devoid of any plastic activity. 
A discreet symmetry breaking is induced by the choice of initially deforming in the forward sense so that the stress free residual strain $\delta\Gamma$ is slightly positive. 
We now define $\Sigma^-$ and $\Sigma^+$  as the negative and positive stress bounds of the elastic range, and $\Delta \Gamma= \Sigma^+ - \Sigma^-$ as its extent. We have immediately:
\begin{equation}
\Sigma^- = -\Gamma + \delta \Gamma \;, \quad
\Sigma^+ = \Gamma + \delta \Gamma \;, \quad
\Delta \Sigma = 2 \Gamma 
\label{isotropic-hardening}    
\end{equation}
Neglecting the slight positive polarization $\delta\Gamma << \Gamma$, for cycling amplitudes $\Gamma < \Gamma_0$ we thus recognize an \emph{isotropic} strain hardening behavior: the elastic stress range $[\Sigma^-, \Sigma^+]$ grows almost perfectly symmetrically upon deformation.

%{It shares similarities with mesoscale models developed for crystalline plasticity~\cite{Trusk-Puglisi-JMPS02,Trusk-Puglisi-CMT02,Trusk-Baggio-PRL19}. In contrast with models of continuum plasticity which require the prescription of an ad-hoc hardening law~\cite{Ohno-JAM82,Chaboche-IJP86,Chaboche-IJP89,Toth-MM00,Wollmershauser-IJF12}, the hardening behavior is here an emergent property. In particular the dissipation transition at $\Gamma_0$ can be interpreted as a transition between an isotropic and a kinematic behavior.}

\begin{figure*}[h!]

\includegraphics[width = .33\textwidth]{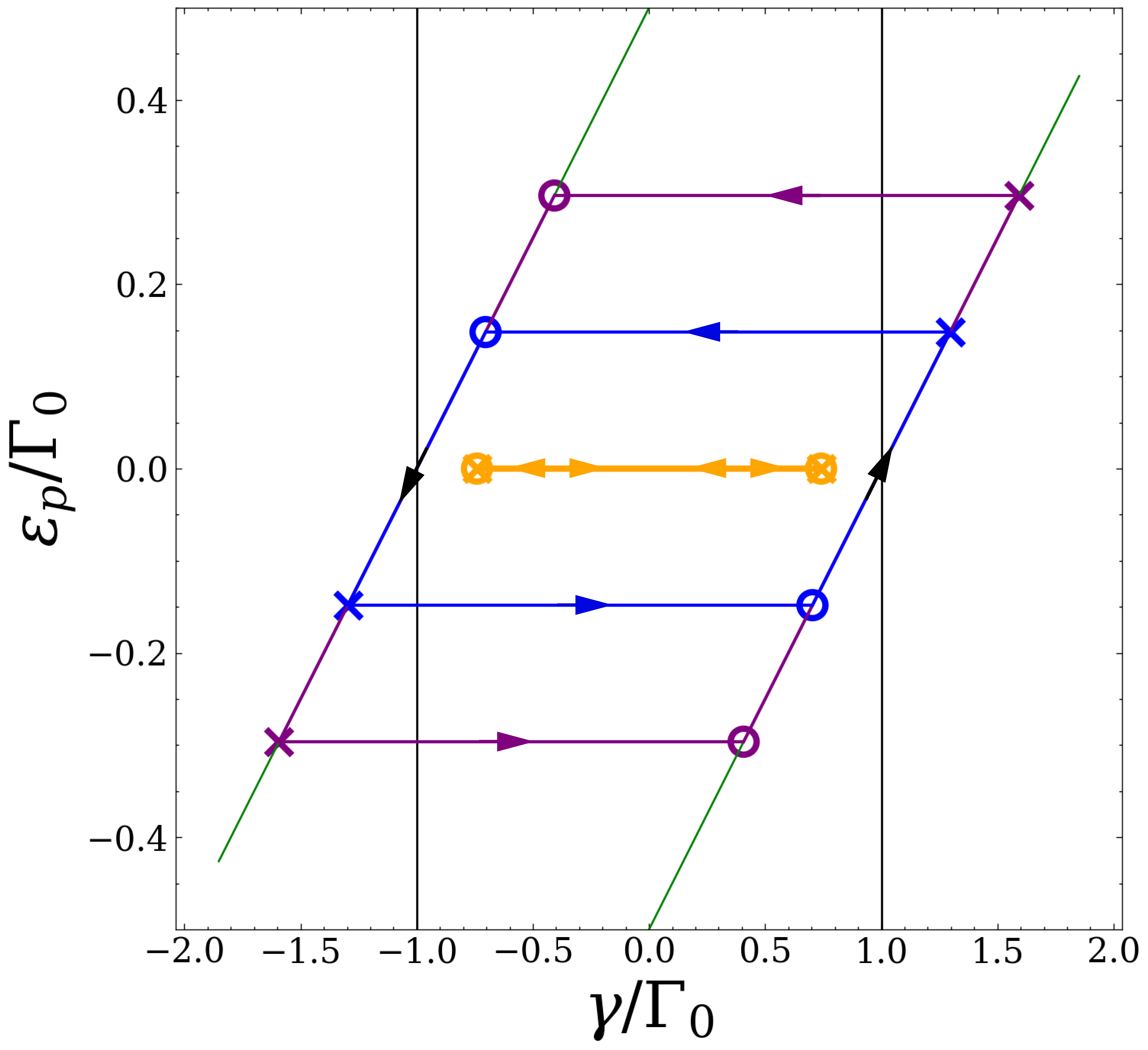}
\includegraphics[width = .33\textwidth]{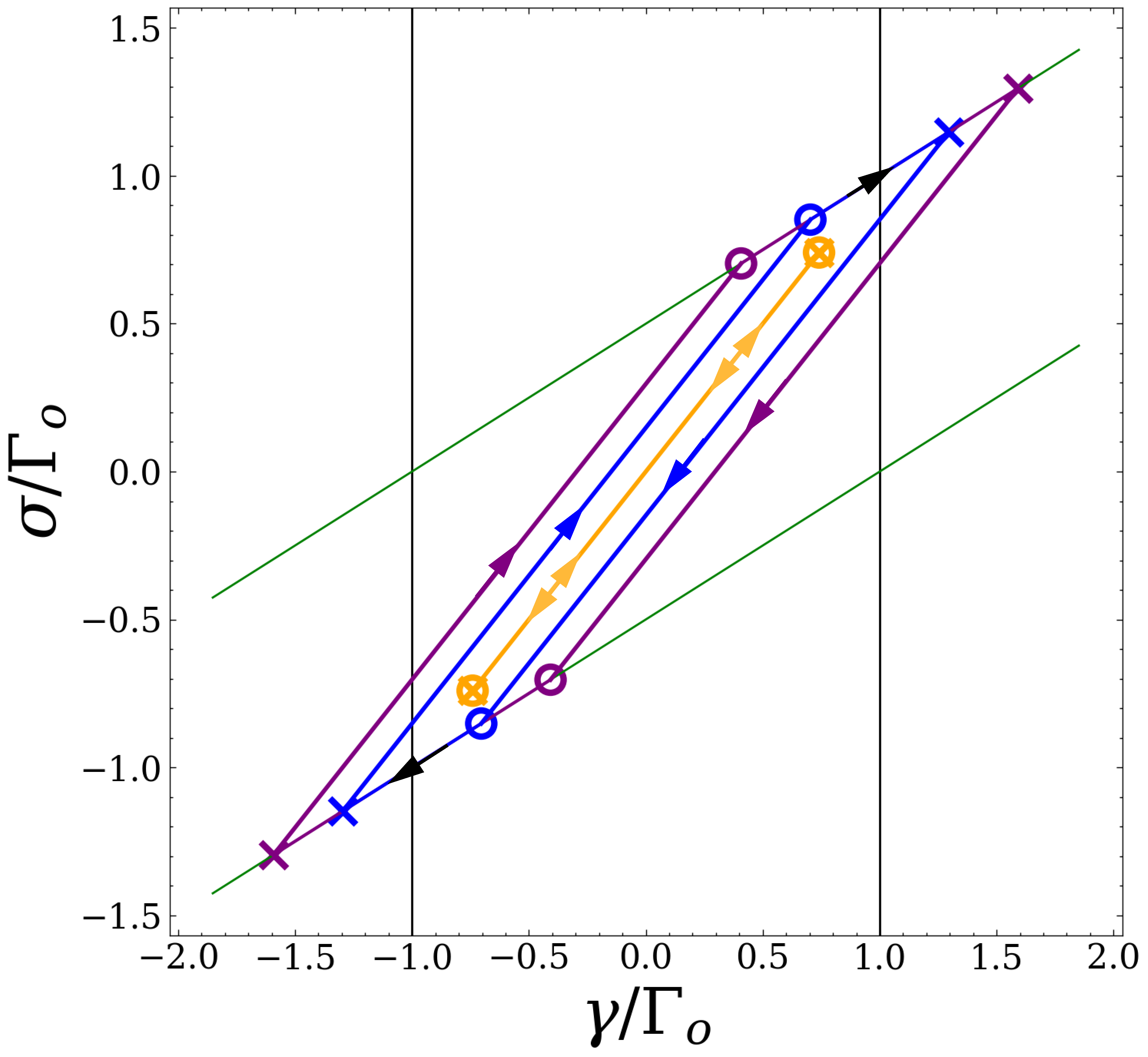}
\includegraphics[width = .33\textwidth]{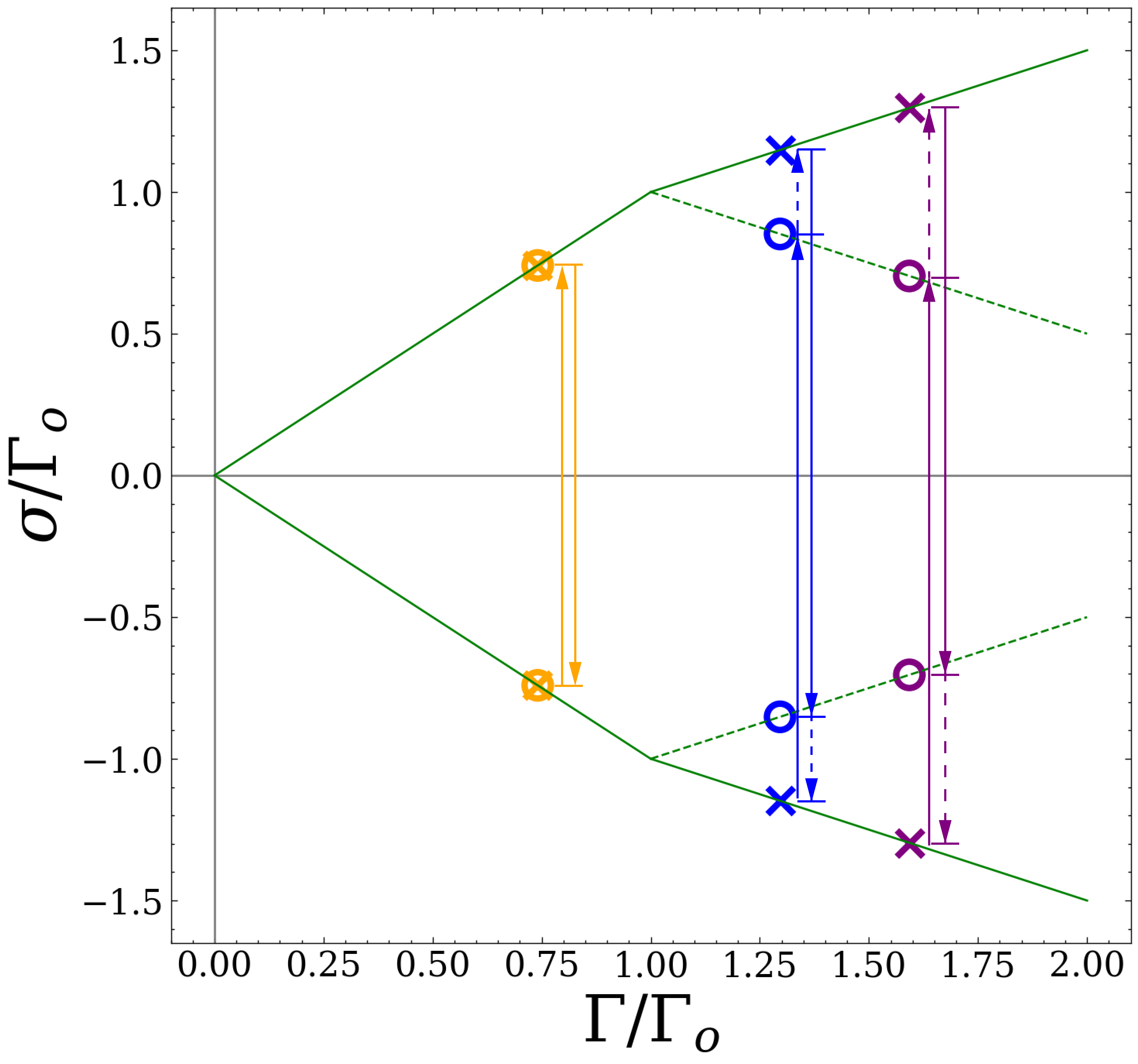}
\caption{Simplified model of hysteresis loops and yield surface.  
a) Plastic strain vs. applied strain at various cycling amplitudes.  
b) Stress vs. applied strain at various cycling amplitudes.
c) Yield surface location, $[\Sigma^-,\Sigma^+]$, in the forward ($[\Sigma_>^-,\Sigma_>^+]$) and reverse ($[\Sigma_<^-,\Sigma_<^+]$) driving states. 
In all plots, crosses indicate the shear reversal points ($\Sigma_>^+, \Sigma_<^-$), and open circles indicate the yield stress values ($\Sigma_>^-, \Sigma_<^+$).
Gold curve and points corresponds to a typical cycling amplitude below the transition at $\Gamma<\Gamma_0$, and the blue and purple curves correspond to two typical cycling amplitudes above the transition at $\Gamma>\Gamma_0$.
Specifically, $\Gamma/\Gamma_0=0.74,1.30,1.59$, respectively for the gold, blue, and purple amplitudes. 
a) Plastic strain, $\epsilon$ vs applied strain $\gamma$.  
Solid green lines indicate the idealized hardening behavior with slope=$\kappa=1/2$ that intersect the horizontal axis at $\gamma/\Gamma_0=\pm1$.
b) Stress, $\sigma$ vs applied strain, $\gamma$.  
Solid green lines indicate the idealized hardening curve with slope=$1-\kappa=1/2$ that contain the points $(\sigma/\Gamma_0,\gamma/\Gamma_0)=(1,1)$ and $(-1,-1)$ respectively.
c) Yield stress, $\Sigma_>^+, \Sigma_<^-$, (dashed green line) and stress at the turning point, $\Sigma_>^-, \Sigma_<^+$, (solid green line) as a function of cycling amplitude, $\Gamma$.
Solid blue and purple lines correspond to elastic range in the forward and reverse shearing direction with the sense of the shearing indicated by the arrow heads.
Dashed blue and purple lines correspond to the post-yield plastic range.
}
\label{fig:hardening}
\end{figure*}

For cycling amplitudes beyond the transition, with $\Gamma > \Gamma_0$, we now get a hysteretic limit cycle. 
We can distinguish between a forward shearing branch and a backward shearing branch, each of which has a pure elastic segment and a plastic segment. 
We define $\Sigma_>^-$ and $\Sigma_>^+$  as the stress bounds of the elastic range in the forward shearing branch and $\Sigma_<^-$ and $\Sigma_<^+$  as the stress bounds of the elastic range in the backward shearing branch. 
A simple way to summarize our main result above on cycling is that the elastic range 
\begin{equation}
\Sigma_<^+ -\Sigma_<^-=\Sigma_>^+ -\Sigma_>^-=2\Gamma_0
\end{equation}
is independent of the cycling amplitude.
This uniform translation of the region in stress space of elastic stability is the hallmark of \emph{kinematic} hardening laws in continuum plasticity theory.
%Likewise we define $\Delta \Gamma_>= \Sigma_>^+ - \Sigma_>^-$ and $\Delta \Gamma_<= \Sigma_<^+ - \Sigma_<^-$  as the extents of the forward and backward elastic ranges, respectively.

To illustrate, we introduce a simplified model of the hardening behavior that neglects certain details present in the data we presented above but retains the essential features.
First, we neglect the symmetry breaking between forward and reverse shearing and assume that all hysteresis loops are symmetric under $\gamma\rightarrow-\gamma$.
Second we suppose that there is absolutely no plastic activity until we reach the yield threshold $\gamma_*$.
Third, we assume that once the system begins to yield plastically, it flows with a constant plastic strain rate, $\kappa=d\epsilon/d\gamma$, and that $\kappa$ is independent of the cycling amplitude.
Finally, we assume that in the hysteretic regime, the length of the elastic segments is given precisely by our argument above so that $\gamma_*=\Gamma_0-(\Gamma-\Gamma_0)$.

Figure~\ref{fig:hardening} shows how these assumptions imply a rather simple behavior.
In figure~\ref{fig:hardening}a, we plot the plastic strain vs. applied strain for a cycle, in figure~\ref{fig:hardening}b, we plot the stress vs. applied strain for a cycle, and in figure~\ref{fig:hardening}c, we plot the location of the yield surface, $\Sigma^+$ and $\Sigma^-$ (i.e. the boundary of the elastic regime) vs. the cycling amplitude for both the forward and reverse loading case.
%In figure~\ref{fig:hardening}, we plot the turning point stresses, $\Sigma_<^+$ and $\Sigma_>^-$, as solid green lines and we plot the yield stresses, $\Sigma_<^-$ and $\Sigma_>^+$ as dashed green lines.
For $\Gamma<\Gamma_0$, the $\epsilon$ vs. $\gamma$ curves  are just flat lines at $\epsilon=0$, while for $\Gamma<\Gamma_0$, the hysteresis loops are perfectly symmetrical trapezoids with slope zero during elastic loading and slope $\kappa$ during the plastic flow as can be seen in figure~\ref{fig:hardening}a.
The amplitude of the hysteresis, $\epsilon_p$, is then perfectly linear in $\Gamma$ with $\epsilon_p=\kappa(\Gamma-\Gamma_0)$
For the stress-strain curves, in figure~\ref{fig:hardening}b, we would then have straight lines of slope $1$ for the elastic segments and lines with slope $1-\kappa$ for the plastic flow.  
%If we translate this into values for the yield stress, $\sigma_*$, and the stress at the turning points, we get...
In figure~\ref{fig:hardening}c, we show the end points of the elastic range $[\Sigma^-,\Sigma^+]$ for both forward and reverse loading.

In the toy model, we set $\kappa=1/2$ to be consistent with the approximate form of the curves shown above in figure~\ref{fig:fig2}.
The constant slope of $1/2$ in the toy model explains the approximate slopes of $-1/2$ in the yield stress vs cycling amplitude curves and the approximate slope of $+1/2$ in the hysteresis amplitude vs cycling amplitude curves in figure~\ref{fig:fig2}. 
%We explain this behavior by taking the derivative of $\sigma=\gamma-\epsilon$ at the onset with respect to $\Gamma$ and requiring that $\frac{d\gamma_*}{d\Gamma}=-1$, from our observation in figure \ref{fig:fig2} (top), then one would get $\frac{d\sigma_*}{d\Gamma}+\frac{d\epsilon_p}{d\Gamma}=-1$. 
%\DV{Message/argument unclear to me}
%Notice that we associate the plastic strain plateau value at the reverse turning point, $\epsilon_p^-$, with the forward $\sigma_*$ and $\gamma_*$ and in that we ignore the creeping regime before the onset and approximate the hysteresis curves in figure \ref{fig:hysteresis} (top) as perfect parallelograms.
%Before $\Gamma_0$, there is no plasticity and $\epsilon_p^{+}=\epsilon_p^{-}>0$.
%This positive residual value is the result of the arbitrary choice to shear the virgin state in the forward direction first.  
Of course, the hysteresis curves in figure~\ref{fig:hysteresis} clearly show that $d\epsilon/d\gamma$ is not constant, but, rather, increases with strain.
Furthermore, this simplified model implies that the hysteresis amplitude is linear in $\Gamma-\Gamma_0$, whereas we showed above that it is a power law in $\Gamma-\Gamma_0$ with an exponent bigger than $1$.
Nevertheless,  for purposes of the present discussion, the main point is that there is a regime below $\Gamma_0$ consistent with the isotropic hardening phenomenology of continuum plasticity models, and a regime above $\Gamma_0$ consistent with the kinematic hardening phenomenology of continuum plasticity models.
In particular, above $\Gamma_0$, the difference between the forward stress at the turning point and the yield stress in the reverse direction is precisely $2\Gamma_0$  -- the length of the solid blue and solid purple lines in figure~\ref{fig:hardening}c, $\Sigma_<^+-\Sigma_<^-$.
This shift, in stress space, of the yield surface is precisely the phenomenology which would result from a kinematic hardening law.
So it is interesting to note that we observe this transition from isotropic to kinematic hardening behavior precisely at the critical cycling amplitude, $\Gamma_0$.

%\section{Summary and conclusion}
\section{Summary and discussion}
\label{sec:conclusion}

In this work, we have explored the behavior of an elasto-plastic automaton model of amorphous systems with varying dimensionless compression modulus, $K/\mu$, both under cyclic shear and steady forward shear.
The ratio $K/\mu$ enters into our model, precisely as in the continuum Eshelby problem, through the ratio $\alpha=2K/(K+\mu)$.
We have shown that, while the magnitude of the elementary macroscopic stress relaxation accompanying a single shear transformation is invariant with respect to $\alpha$ (assuming that the local yielding condition does not depend on $K/\mu$), the stress \emph{redistribution} accompanying that shear transformation in all space including at the yielding site itself, is precisely proportional to $\alpha$.
This has profound implications for the emergent behavior.  
In particular as $K/\mu$ increases, and $\alpha$ increases monotonically toward $2$, the Eshelby back stress, $\sigma_0$, becomes increasingly negative.
$\sigma_0$, plays a similar role in determining the location of the elastic to reversible-plastic transition in cyclic shearing at a cycling amplitude of $\Gamma_0$ as it does in determining the flow stress, $\sigma_y$, in forward steady shear.
However, the dependence on $\sigma_0$ is quantitatively different for $\Gamma_0$ than for $\sigma_y$, and we were able to explain these quantitative differences in terms of the forms of the $P(\sigma)$ distributions of the local stresses.

In cyclic shearing, for the incompressible case, we have shown in a previous study \cite{elgailani2022anomalous} that the critical amplitude, $\Gamma_0$, is given by $(1+\sigma_0)/2$, and that for any cycling amplitude with $\Gamma>\Gamma_0$, we have a relatively sharp onset of plasticity at $\gamma_*=\Gamma_0-(\Gamma-\Gamma_0)$.
The plasticity onset is quite sharp at the critical amplitude, $\Gamma_0$, and becomes less sharp at larger $\Gamma$ with a creeping regime with low plastic strain rate emerging at $\gamma<\gamma_*$.
Here, we found the same results for the compressible case after adjusting $\sigma_0$ to account for the finite compressibility.
%Here, we have found that $\Gamma_0$ for the compressible case is also given by the same relation. 
This means that $\Gamma_0$ decreases with increasing $K/\mu$ as $\sigma_0$ becomes increasingly negative.
%After accounting for this dependence of $\Gamma_0$ on $K/\mu$, we found that the previous results in \cite{elgailani2022anomalous} for cyclic shearing in the incompressible case continue to hold.
The only other adjustment we needed to make here for the compressible case was to scale the plastic strain amplitude, $\epsilon_p$, by $\Gamma_0$ in order to obtain a collapse for $\epsilon_p$ vs. $\Gamma$ for different $K/\mu$.
The other observations, including the form of the $P(\sigma)$ distributions, go through essentially unchanged.

We have also shown that the shape of the $\epsilon$ vs. $\gamma$ loading curves near the onset at $\gamma_*$ can be reasonably well inferred from the $P(\sigma)$ distributions.
In particular, the rounding of the loading curves and the creep regime at $\gamma<\gamma_*$ at larger values of $\Gamma-\Gamma_0$ further away from the critical point, were well captured by $P(\sigma)$.
The plastic strain inferred from $P(\sigma)$ overestimated, slightly but systematically, the actual values.
However, this did not seem to affect the ability to capture the rounding of the loading curves at high $\Gamma-\Gamma_0$.
This gives us hope that mean-field approaches based on analysis of the stress distribution without consideration of the spatial structure of the stress field should be able to capture the onset at $\gamma_*$ and its rounding with the emergence of a creeping regime below $\gamma_*$.

% As in the case of compressible systems, we have found that: i) increasing the the cycling amplitude, $\Gamma$, beyond $\Gamma_0$, the systems become anomalously softer--exhibit plasticity at lower strains-- despite the fact that their energy continues to go down.
% % ii) If we characterize the ER->RPR transition by the plastic strain amplitude, the transition has a second-order nature where as the cycling amplitude increases, the amplitude of plasticity contiguously increases from precisely zero below the transition point at $\Gamma_0$ to a finite value above $\Gamma_0$.
% ii) The transition from ER to RPR is characterized by a power law onset of the plastic strain amplitude with $\epsilon_p=1.4(\Gamma-\Gamma_0)^{1.2}$.
% iii) The plastic strain rate for systems in the RPR jumps from virtually zero below the plasticity onset strain at $\gamma_*$ to a finite value.
% The magnitude of the jump is essentially insensitive to the strain amplitude or $K/\mu$.
% iv) The local stress distribution,$P(\sigma)$, for all $K/\mu$, has a pronounced shoulder at $\sigma_0$ that links $\Gamma_0$ and $\gamma_*$ to $\sigma_0$; $\Gamma_0=(1+\sigma_0)/2$ and $\gamma_*=\Gamma_0-(\Gamma-\Gamma_0)$. 
% v) We can obtain a very good approximation of the shape of the loading curves from $P(\sigma)$.

In the case of forward steady shear, the loading curves for different $K/\mu$ collapse with remarkable precision after rescaling the stress and strain.
At early stages for smaller plastic strains, we found a good collapse after rescaling the plastic strain by $\alpha$, indicating that the incurred plastic strain per unit applied strain is proportional to $\alpha$. 
At later stages, on approach to yielding, we used a slightly different scaling where we scaled the stress by the flow stress, $\sigma_y$, and we used the scaling properties of $P(\sigma)$ to explain how $\sigma_y$ depends on $K/\mu$ via $\sigma_0$.
Similar to the case of cyclic shear, we found that $P(\sigma)$ has a pronounced shoulder at $\sigma_0$.
%However, the dependence of $P(\sigma)$ on $\sigma_0$ is  different for forward shear than for cyclic shear.
%In forward shear, the $P(\sigma)$ distributions for different $K/\mu$ collapse almost perfectly if we rescale $P(\sigma)$ by $(1-\sigma_0)$ and shift the local stresses such that the shoulders all sit at $\sigma_0$.
In forward shear, when we transformed to a shifted and scaled value of the stress, $\sigma^\prime=(\sigma-\sigma_0)/(1-\sigma_0)$, the upper limit of $P(\sigma^\prime)$ remained at $1$, while the shoulder at $\sigma_0$ was mapped to $\sigma^\prime=0$ for all $K/\mu$.
After the transformation, the distributions collapsed almost perfectly.
This implied a simple expression for $\sigma_y$: $\sigma_y=\sigma_y^\prime+(1-\sigma_y^\prime)\sigma_0$
where $\sigma_y^\prime$ is the average of the rescaled universal $P(\sigma^\prime)$ distribution which is just a constant.
Therefore, as in the case of $\Gamma_0$, $\sigma_y$ also decreases with $K/\mu$ but in a quantitatively different way.

The model we employed here is based on a spatially uniform and fixed-in-time local strain energy function where every site in the system has the same microscopic threshold for yielding at a local elastic strain of unity.
A mesoscopic tensorial model (MTM) of crystalline plasticity relying on the construction of an energy density respecting the global symmetry of the Bravais lattice has recently been developed in a close vein~\cite{Trusk-Baggio-PRL19,Baggio-EJMA23}.
Other so-called elasto-plastic models which have been employed in the literature~\cite{nicolas2018deformation,C4SM00395K,ISI:000289524200004,PhysRevLett.89.195506,Talamali:2012aa,doi:10.1063/5.0102669,ISI:000391865100010,ISI:000328694700010,Budrikis:2017aa,ISI:000447093900001,Khirallah-PRL2021,ISI:000491996300004,ISI:000446138000009,ISI:000378873300008,Lin:2016aa,Liu:2021un} make varying assumptions about the microscopic yielding conditions and the consequences -- and temporal dynamics -- of a yielding event. 
One study by Talamali et. al.~\cite{Talamali:2012aa} used a stochastic choice for the plastic strain increment chosen randomly from a uniform distribution with a tunable upper bound.
In that study, a similar decrease in the emergent, macroscopic yield stress was observed when the characteristic strain increment was increased as we saw here when increasing $\alpha$.
The motivation in that study to vary the characteristic amplitude of the plastic strain was to make connections to elastic depinning phenomena. 
In the present study, we were, rather, motivated to study the consequences of changing dimensionless compressibility, and we were
able to precisely rationalize and explain the behavior of both $\Gamma_0$ and $\sigma_y$, directly in terms of the $\alpha$ parameter and its affects on the  $P(\sigma)$ distribution via the Eshelby stress. 

Shang et. al.~\cite{Shang:2020aa} have studied the avalanche behavior in an MD simulation focusing on the transient regime before reaching steady plastic flow.
They prepared systems with different quenching protocols resulting in different $K/\mu$ and found a pronounced effect of $K/\mu$ on the size dependence of the avalanches.
%Furthermore, they showed a strong dependence on $K/\mu$ for the macroscopic stress vs. plastic strain with precisely the same effect we see here: higher stress at given plastic strain for lower $K\mu$.
Furthermore, they showed a strong dependence on $K/\mu$ for the macroscopic stress vs. plastic strain with higher stress at given plastic strain for lower $K/\mu$.
However, they did not explore the possible connection between this effect and the amplitude of stress redistribution.
It would be interesting to explore this connection further in particle-based models like MD and non-linear elasto-plastic models where the elastic properties do depend on sample age and preparation where proximity to local yielding thresholds generally results in elastically soft spots which become less prevalent in more slowly quenched or annealed systems.
%Although, the (discontinuous) stress-strain curves in our present \emph{linear} elasto-plastic model may depend on sample preparation and loading history, the elastic branches are all perfectly linear with the elastic moduli entering simply as model parameters.

The results we presented here follow from modeling amorphous systems in what we believe to be the simplest possible way with no free parameters other than $K/\mu$.
%Nonetheless, we believe that the main results, such as the anomalous softness and the dependency of $\Gamma_0$ on compressibility, will qualitatively hold in experiments and more realistic simulations.
Nonetheless, we believe that the main results, namely the dependence of $\Gamma_0$ and $\sigma_y$ on compressibility, will qualitatively hold in experiments and particle-based simulations (such as molecular dynamics) where local elastic properties and yielding thresholds may vary in space and be distributed more broadly.
The main underlying ingredient is the increased amplitude of stress \emph{redistribution} after a yielding event in systems which are less compressible.
It would be reasonable to assume that in more realistic models and in experiments, an \emph{effective} $\alpha$ would emerge to set the typical amplitude of stress redistribution.
Our findings should motivate future experiments and simulations to have a more precise quantification of the role of compressibility in emergent plastic yielding behavior in cyclic shear and monotonic forward shearing.

It is important to note, that we implicitly work in units of stress and strain that are determined by the elementary yielding strain threshold (conventionally set here to unity).
In more realistic models such as MD simulations or in experiments, it would be important to disentangle the effects of changes in \emph{elementary} threshold strains from the Poisson ratio effects we describe here.
A simple change in the elementary yield threshold would result in a trivial scaling of the stresses.
However, we expect one would see the Poisson ratio effects we see here after appropriately scaling stresses and strains to account for any shifts in the elementary microscopic thresholds.

In mean-field theories of amorphous matter, ~\cite{lemaitre2007plastic,Agoritsas:2015aa,C6SM02702D,Ferrero:2019aa, doi:10.1063/5.0033196,PhysRevLett.128.198001,Ferrero_2021,PhysRevLett.126.255501,PhysRevLett.127.248002,cochran2022slow,PhysRevMaterials.6.065601}, the amplitude of the stress redistribution, along with the distribution from which the stress changes are drawn, plays an important role.
In the mean-field approaches, and even other elasto-plastic models constructed in real-space, the amplitude is usually taken as a purely ad-hoc parameter and rarely motivated physically.
Here, we show precisely how the amplitude arises from the form of the elastic interactions and how it depends on the elastic moduli.
In the model we study here, the yield condition contains no dependence on pressure or  any other field representing any sort of free volume.
These dependencies are often invoked in the materials science community~\cite{Sun:2016we,Schuh:2007wv} to explain the relation between Poisson ratio and ductility.
So it is particularly striking that, despite the simplicity of our model, we still see such a strong dependence of the plasticity on the Poisson ratio.
%It is particularly striking that, despite this, we see such a strong signature of the compressibility even in our simple model which does not allow for any non-linear volumetric changes like dilation or compaction, nor include any explicit dependence on volumetric strain in the yielding condition -- or introduce any other field accounting for effects of any sort of free volume.
%In the future it would be interesting to allow for permanent dilation/compaction and include pressure dependence in the yield condition.  

Naturally, there are many more questions remaining to be addressed in future studies. 
i) Here all systems were prepared in an initial rapid quench.  The issue of thermal annealing, or zero temperature aging~\cite{doi:10.1063/5.0102669}, of the samples before subjecting them to shearing should be investigated.
ii) We have used a simple single-amplitude cycling protocol.  More complex loading protocols: e.g. small amplitude cycling to anneal mechanically followed by large amplitude cycling to probe yield or more complex protocols would be useful to study.
%ii) In our initialization protocol, we impose a random plastic field of $\pm2$ at any site and run the automaton dynamics once. It would be interesting to repeat the study with a different initialization protocol such as imposing a random integer multiple of $\pm2$ at the sites or repeating our initialization process multiple times.
iii) In previous work \cite{Khirallah-PRL2021}, some of us studied the transition to \emph{irreversible}, diffusive, cyclic shear at a larger transition amplitude, $\Gamma_y$, for incompressible systems. 
A similar study for compressible systems is in order.  
One might wonder if $\Gamma_y$ in cycling is equal to $\sigma_y$ in forward steady shear, or at least has a similar dependence on $\alpha$.
iv) 
The effect of finite compressibility on the nature of avalanches, both in cycling and steady shear, should be investigated.
Properties such as their size distribution, spatial correlations in real space, and finite size effects would be of interest.
%Of course, we expect universal properties such as scaling exponents to be unchanged, but non-universal properties such as the form of scaling functions and prefactors in scaling laws would likely have a dependence on $K/\mu$ via $\alpha$.
v)  More realistic models with distributions of thresholds and plastic strain increments should be studied~\cite{doi:10.1063/5.0102669}.

\clearpage
\begin{acknowledgments}
We acknowledge computing support on the Discovery computing cluster at the MGHPCC.
\end{acknowledgments}

% \section{Appendix}
% The deformation modes we refer to as modes 1, 2, and 3 correspond to the three components of the total strain tensor in Voigt notation: 
% \begin{eqnarray}
% \gamma_1&=&\gamma_{xx}-\gamma_{yy}\\
% \gamma_2&=&\gamma_{xy}+\gamma_{yx}\\
% \gamma_3&=&\gamma_{xx}+\gamma_{yy}
% \end{eqnarray}
% And their Cauchy stress conjugates:
% \begin{eqnarray}
% \sigma_1&=&\frac{\sigma_{xx}-\sigma_{yy}}{2}\\
% \sigma_2&=&\frac{\sigma_{xy}+\sigma_{yx}}{2}\\
% \sigma_3&=&\frac{\sigma_{xx}+\sigma_{yy}}{2}
% \end{eqnarray}
%  In this study, we shear the system in mode $2$ so we are mainly interested in its corresponding quantities.

\bibliography{Bib.bib}

\clearpage
\onecolumngrid
\begin{center}
\rule{0.8\linewidth}{0.4pt}\\[8pt]
{\large\textbf{Supplemental Material}}\\[4pt]
{\normalsize Finite compressibility and strain hardening in elasto-plastic models of amorphous matter}\\[8pt]
\rule{0.8\linewidth}{0.4pt}
\end{center}

\setcounter{equation}{0}
\setcounter{figure}{0}
\setcounter{table}{0}
\renewcommand{\theequation}{S\arabic{equation}}
\renewcommand{\thefigure}{S\arabic{figure}}
\renewcommand{\thetable}{S\arabic{table}}

\section*{Staggered grid discretization of Eshelby's problem}

The automaton model we use here is almost identical to the one employed in reference~\cite{Khirallah-PRL2021}.
However, we have changed the details of how the elasticity is discretized.
In the present study have used a so-called staggered grid discretization which is a common procedure in seismology~\cite{10.1785/BSSA0660030639,Levander:1988ue,Randall1991MultipoleBA,Virieux:1986tb,10.1785/BSSA0860041091} and has recently started to see more use in the materials science and theoretical solid mechanics communities~\cite{Schneider:2016tn}.
The discretization used here results in precisely the same $K_{11}$ and $K_{22}$ (defined below) as in reference~\cite{Khirallah-PRL2021}, but different $K_{12}$.
The $K_{12}$ used in reference~\cite{Khirallah-PRL2021} had some pathological behavior which is not present in the staggered grid discretization.
Although we do not think these pathologies affected the results in that study or would present any problems in the present study, we nevertheless have opted to use the staggered grid technique here.

We start by defining a displacement field $u_\alpha[I,J]$ on a regular grid with co-ordinates indices $I,J$.
The grid is square with $0\leq I <L$, $0\leq J < L$.
We define the total strain, $\epsilon$, from the displacement field using the staggered grid technique~\cite{Schneider:2016tn} and assuming periodic boundary conditions on $u$ so that $u[I+mL,J+nL]=u[I,J]$ for any integers $m,n$.
This gives:
\begin{equation}
\gamma_{xx}[I,J]=u_x[I+1,J]-u_x[I,J]
\end{equation}
\begin{equation}
\gamma_{xy}[I,J]=u_y[I,J]-u_y[I-1,J]
\end{equation}
\begin{equation}
\gamma_{yx}[I,J]=u_x[I,J]-u_x[I,J-1]
\end{equation}
\begin{equation}
\gamma_{yy}[I,J]=u_y[I,J+1]-u_y[I,J]
\end{equation}
For a physical interpretation of this staggered difference scheme, see reference~\cite{Schneider:2016tn}
Note that not all $\gamma$ fields are derivable from a $u$ field.
The $\gamma$ fields which \emph{are} derived from a $u$ field this way are kinematically compatible within our discretization scheme.
We can define a local elastic strain energy as
\begin{equation}
\phi[I,J]=\frac{\mu}{2}\left((\gamma_1[I,J]-\epsilon_1[I,J])^2+(\gamma_2[I,J]-\epsilon_2[I,J])^2\right)+\frac{K}{2}(\gamma_3[I,J])^2
\end{equation}
where $\mu$ and $K$ are the shear and compression moduli, $\gamma_1$ and $\gamma_2$ are the two shear components and $\gamma_3$ the dilatant component of the total strain:
\begin{equation}
\gamma_1=\gamma_{xx}-\gamma_{yy}
\end{equation}
\begin{equation}
\gamma_2=\gamma_{xy}+\gamma_{yx}
\end{equation}
\begin{equation}
\gamma_3=\gamma_{xx}+\gamma_{yy}
\end{equation}
and $\epsilon_1$ and $\epsilon_2$ are two prescribed plastic strain fields or so-called eigenstrains.
The total elastic strain energy is then simply a sum over the lattice.
\begin{equation}
\phi=\sum_{IJ}\phi[I,J]
\end{equation}
The discretized Eshelby problem is then to find a displacement field and the associated strain field which minimizes $\phi$ subject to the prescribed $\epsilon_1$ and $\epsilon_2$.
Note that posing the minimization problem in terms of displacements automatically ensures the resulting strain field is kinematically compatible.
If the prescribed $\epsilon$ field happens to be compatible, then the solution will be trivial with $\gamma=\epsilon$, but we are typically interested in incompatible $\epsilon$ fields.
Because of the linearity of the problem, we can express the solution for any arbitrary prescribed $\epsilon_1$ and $\epsilon_2$ fields in terms of the solution to a unit delta source, 
\begin{equation}
\epsilon_1[I,J]=\delta_{I,0}\delta_{J,0}
\end{equation}
and
\begin{equation}
\epsilon_2[I,J]=\delta_{I,0}\delta_{J,0}.
\end{equation}
Note that these delta source fields are not kinematically compatible and could not have been derived from any displacement field.
We define the strain fields derived from the displacements resulting from the energy minimization problem with a delta source as $K_{11}, K_{12}, K_{21}, K_{22}$ where: $K_{11}$ is the $\gamma_1$ strain resulting from an imposed $\epsilon_1$ eigenstrain,  $K_{12}$ is the $\gamma_1$ strain resulting from an imposed $\epsilon_2$ eigenstrain, $K_{21}$ is the $\gamma_2$ strain resulting from an imposed $\epsilon_1$ eigenstrain, and $K_{22}$ is the $\gamma_2$ strain resulting from an imposed $\epsilon_2$ eigenstrain.
%$K_{12}=K_{21}$ because of a standard Maxwell-type reciprocity relation.      
We specialize here to the incompressible limit where $K/\mu\rightarrow\infty$.
The solution to this energy minimization problem is most easily expressed in Fourier space.
\begin{eqnarray}
\tilde{K}_{11}[p,q]&=&-\frac{4q_{x-}q_{x+}q_{y-}q_{y+}}{\mathcal{D}}=-\frac{4\Delta^2_x\Delta^2_y}{\mathcal{D}}\\
\tilde{K}_{22}[p,q]&=&-\frac{(q_{x-}q_{x+}-q_{y-}q_{y+})^2}{\mathcal{D}}=-\frac{(\Delta^2_x-\Delta^2_y)^2}{\mathcal{D}}\\
\tilde{K}_{12}[p,q]&=&\frac{2 q_{x+} q_{y+} (q_{x-} q_{x+} - q_{y-} q_{y+})}{\mathcal{D}}=\frac{2 q_{x+} q_{y+} (\Delta^2_x-\Delta^2_y)}{\mathcal{D}}\\
\tilde{K}_{21}[p,q]&=&\frac{2 q_{x-} q_{y-} (q_{x-} q_{x+} - q_{y-} q_{y+})}{\mathcal{D}}=\frac{2 q_{x-} q_{y-} (\Delta^2_x-\Delta^2_y)}{\mathcal{D}}
\end{eqnarray}
where
\begin{eqnarray}
\Delta^2_x&=&q_{x+}q_{x-}\\
\Delta^2_y&=&q_{y+}q_{y-}
\end{eqnarray}
are the Fourier transforms of the second ordered centered difference operators in the $x$ and $y$ direction and where
\begin{equation}
\mathcal{D}=(\Delta_x^2+\Delta_y^2)^2
\end{equation}
is the Fourier transform of the the graph Laplacian of the graph Laplacian, 
and where 
\begin{eqnarray}
q_{x+}[p,q]&=&-i (+\exp[+2\pi i p/L]-1)\\
q_{x-}[p,q]&=&-i (-\exp[-2\pi i p/L]+1)\\
q_{y+}[p,q]&=&-i (+\exp[+2\pi i q/L]-1)\\
q_{y-}[p,q]&=&-i (-\exp[-2\pi i q/L]+1)\\
\end{eqnarray}
are the Fourier transforms of the forward and backward difference operators in the $x$ and $y$ directions.
We have used the discrete Fourier transform conventions:
\begin{equation}
\tilde{K}[p,q]=\left(\frac{1}{\sqrt{L}}\right)^2\sum_{I,J}K[I,J]\exp\left[+2\pi (Ip+Jq)/L \right]
\end{equation}
\begin{equation}
K[I,J]=\left(\frac{1}{\sqrt{L}}\right)^2\sum_{p,q}\tilde{K}[p,q]\exp\left[-2\pi (Ip+Jq)/L \right]
\end{equation}
With the solution for the delta sources, we can then write the solution for an arbitrary source field in terms of a convolution:
\begin{equation}
\gamma_{1}[I,J]=\sum_{IJ}K_{11}[M-I,N-J]\epsilon_{1}[M,N] + K_{12}[M-I,N-J]\epsilon_{2}[M,N]
\end{equation}
\begin{equation}
\gamma_{2}[I,J]=\sum_{IJ} K_{21}[M-I,N-J]\epsilon_{1}[M,N] + K_{22}[M-I,N-J]\epsilon_{2}[M,N]
\end{equation}
Then, of course, because of the convolution theorem, we have a decoupled mode-wise relation in for each $p,q$ mode in Fourier space,
\begin{equation}
\tilde{\gamma}_{1}[p,q]=\tilde{K}_{11}[p,q]\tilde{\epsilon}_{1}[p,q] + \tilde{K}_{12}[p,q]\tilde{\epsilon}_{2}[p,q]
\end{equation}
\begin{equation}
\tilde{\gamma}_{2}[p,q]=\tilde{K}_{21}[p,q]\tilde{\epsilon}_{1}[p,q] + \tilde{K}_{22}[p,q]\tilde{\epsilon}_{2}[p,q]
\end{equation}
Our expressions for $K$ were obtained by differentiating the energy with respect to $u_x$ and $u_y$, solving the linear equations for the $u$ fields which give zero energy derivative and then re-inserting those $u$ fields back into the staggered difference scheme to find $\gamma$, and we have not shown those steps here as they are straightforward but tedious.
We note that the present $\tilde{K}_{11}$ and $\tilde{K}_{22}$ are \emph{precisely} the same as in reference~\cite{Khirallah-PRL2021} and as in the continuum Eshelby problem~\cite{picard2004elastic}.
It is only $\tilde{K}_{12}$ and $\tilde{K}_{21}$ which are different in the present staggered-difference scheme. 
The definition of the automaton from a piece-wise quadratic strain-energy function and the initialization procedure are then precisely as in reference~\cite{Khirallah-PRL2021} so we do not elaborate further here. 

\end{document}